\documentclass[a4paper,titlepage,12pt]{article}
\pdfoutput=1
\usepackage{graphicx} 
\usepackage[utf8]{inputenc}
\usepackage{geometry}
\usepackage[english]{babel}
\usepackage{amsmath, amsfonts, graphicx}
\usepackage{mathtools}
\usepackage{graphicx}
\usepackage{float}
\usepackage{mciteplus}
\usepackage{bbold}
\usepackage{pstricks}
\usepackage{xcolor}
\usepackage[labelfont=bf,font={small}]{caption}
\usepackage{pgfplots}
\usepackage{empheq}
\usepackage{tikz}
\usetikzlibrary{positioning, arrows}
\usetikzlibrary{external}
\usepackage{footmisc}
\usepackage{multirow}
\usepackage{url}
\usepackage[breakable, theorems, skins]{tcolorbox}
\usepackage{enumerate}
\usepackage{physics}
\usepackage{bm}
\usepackage{braket}
\usepackage{epstopdf}
\usepackage{hyperref}
\usepackage[OT2,T1]{fontenc}
\DeclareSymbolFont{cyrletters}{OT2}{wncyr}{m}{n}
\DeclareMathSymbol{\Sha}{\mathalpha}{cyrletters}{"58}
\date{June 2024}

\definecolor{darkblue}{RGB}{0,0,139}

\hypersetup{
    colorlinks,
    citecolor=blue,
    filecolor=blue,
    linkcolor=blue,
    urlcolor=blue
}

	\numberwithin{equation}{section}
\begin{document}

\begin{titlepage}

\setcounter{page}{1} \baselineskip=15.5pt \thispagestyle{empty}

\vfil

${}$
\vspace{1cm}

\begin{center}

\def\thefootnote{\fnsymbol{footnote}}
\begin{changemargin}{0.05cm}{0.05cm} 

\begin{center}

{\Large \bf Probing the singularity at the holographic screen via $q$-holography} \\

\end{center}
\end{changemargin}

~\\[1cm]
{Andreas Belaey,\footnote{{\protect\path{andreas.belaey@ugent.be}}} Thomas G. Mertens,\footnote{{\protect\path{thomas.mertens@ugent.be}}} Jacopo Papalini,\footnote{{\protect\path{jacopo.papalini@ugent.be}}}}
\\[0.3cm]
{\normalsize { \sl Department of Physics and Astronomy
\\[1.0mm]
Ghent University, Krijgslaan, 281-S9, 9000 Gent, Belgium}}\\[3mm]

\end{center}

 \vspace{0.2cm}
\begin{changemargin}{01cm}{1cm} 
{\small  \noindent 
\begin{center} 
\textbf{Abstract}
\end{center} 
We study the emergence of $q$-deformed spacetime in a lower-dimensional gravitational system whose asymptotic region geometrizes the global symmetry of a $q$-deformed CFT. More precisely, we consider the 2d sinh dilaton gravity model, whose classical metric solutions exhibit a curvature singularity at the holographic boundary. Our aim is to probe this UV near-boundary regime by injecting a small probe into the bulk, and identifying the geometrical features it observes. At the level of the two-point correlator, we see the emergence of $q$-deformed hyperbolic disk isometries. By formulating DSSYK in terms of the analogous sine dilaton gravity model, we expect this $q$-deformed holographic duality to persist.
}
\end{changemargin}
 \vspace{0.3cm}
\vfil
\begin{flushleft}
\today
\end{flushleft}

\end{titlepage}

\tableofcontents

\setcounter{footnote}{0}

\section{Introduction}

Going beyond the conventional AdS/CFT paradigm \cite{Maldacena:1997re} is of considerable interest; primarily because we do not live in an AdS universe, but also because it allows us to distill lessons on holography in the more general context. Within the context of holography and string theory, some important examples have been e.g. \cite{Giveon:1999px,Klebanov:2000hb,Kanitscheider:2008kd,Taylor:2015glc}. To achieve further progress towards this goal, especially when aiming for quantized gravity, exactly solvable models serve as powerful benchmarks. In this light, lower-dimensional dilaton gravity models have attracted significant attention in the past years. This was sparked by studies on Jackiw-Teitelboim (JT) gravity \cite{Jackiw:1984je, Teitelboim:1983ux,Almheiri:2014cka, Jensen:2016pah, Maldacena:2016upp, Engelsoy:2016xyb, Cotler:2016fpe, Stanford:2017thb, Kitaev:2018wpr, Mertens:2017mtv, Mertens:2018fds, Lam:2018pvp, Harlow:2018tqv}, its direct relation with microscopic SYK models \cite{kitaev2015talk1,kitaev2015talk2, Sachdev:1992fk}, its lessons on spacetime wormhole physics \cite{Saad:2019lba}, and its universality for describing the black hole throat for a near-extremal higher-dimensional black hole (see e.g. \cite{Nayak:2018qej,Iliesiu:2020qvm}). JT gravity is but one example of a solvable 2d dilaton gravity model. Indeed, the 2d sinh dilaton gravity model was discovered to be structurally very promising as well, since it is merely the old Liouville gravity models after a field redefinition \cite{Kyono:2017pxs,Mertens:2020hbs,Mertens:2020pfe,Fan:2021bwt,Suzuki:2021zbe,Collier:2023cyw}. Relatedly, within the double-scaling regime of the original SYK model \cite{Berkooz:2018qkz,Berkooz:2018jqr,Lin:2022rbf,Susskind:2022bia,Bhattacharjee:2022ave,Blommaert:2023opb,Blommaert:2023wad,Susskind:2023hnj,Mukhametzhanov:2023tcg,Berkooz:2023cqc,Okuyama:2023bch,Lin:2022nss,Berkooz:2022mfk,Goel:2023svz,Narovlansky:2023lfz,Verlinde:2024zrh,Berkooz:2024evs,Lin:2023trc,Verlinde:2024znh,Almheiri:2024ayc,Almheiri:2024xtw,Bossi:2024ffa,Xu:2024hoc,Xu:2024gfm,Heller:2024ldz,Aguilar-Gutierrez:2024nau,Tietto:2025oxn,Aguilar-Gutierrez:2025hty,Berkooz:2025ydg}, the sine dilaton gravity model has been proposed to describe a holographic dual in \cite{Blommaert:2024ymv,Blommaert:2024whf,Blommaert:2025avl}.\footnote{Sine dilaton gravity has been defined and studied in \cite{Collier:2024kmo} in a different quantization scheme, leading to a continuum model instead which was utilized as a toy model for cosmology in \cite{Collier:2025lux}.} Both of these models have in common that the holographic boundary is not of the conventional asymptotic AdS type. In fact, for sinh dilaton gravity, one readily shows (as we will review later on), that the Ricci scalar of the bulk geometry actually blows up at the holographic screen!

\begin{figure}[h]
    \centering
    \includegraphics[width=0.6\linewidth]{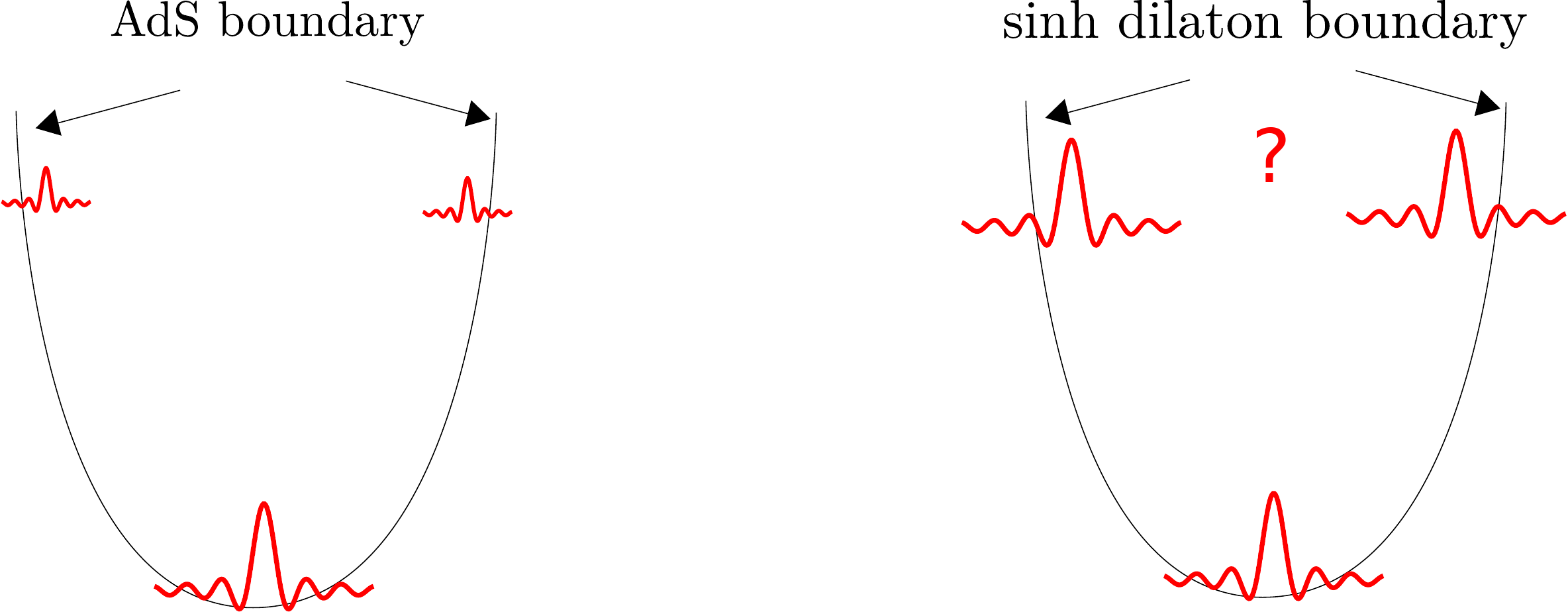}
    \caption{Left: Visualization of quantum gravitational fluctuations being suppressed close to the holographic boundary in an AdS box. Right: Fluctuations are not obviously suppressed for the case where the near-boundary region contains a classical curvature divergence.}
    \label{FigAdSfluct}
\end{figure}

The fact that this model has a curvature singularity exactly where one wants to position the holographic screen presents us simultaneously with a puzzle and an opportunity. The puzzle here is that the holographic boundary is usually at a location in spacetime where gravitational fluctuations are negligible. However, a classical curvature singularity does not seem like a stable anchoring region where one can trust classical gravity. From this, one can expect that there can be a difference between on the one hand the near-boundary limit of the model, and on the other hand the semi-classical $\hbar \to 0$ limit where we turn off gravity quantum fluctuations. In ordinary AdS/CFT, these two regimes give the same answer close to the boundary. But here we will have to be more careful, see Fig.~\ref{FigAdSfluct} for a visualization of the general puzzle studied in this work.

It is of considerable interest currently to have a better understanding of quantum gravity in the case where we do \emph{not} have a boundary where gravitational fluctuations are suppressed, such as happens in cosmological models along the lines of \cite{Witten:2021unn,Chandrasekaran:2022cip,Jensen:2023yxy} and \cite{Susskind:2021dfc,Shaghoulian:2021cef,Shaghoulian:2022fop,Franken:2023pni}. Hence, one can view our current investigation as following a similar goal.

Generically, studying singularities in quantum gravity is hard, since the cosmic censorship hypothesis teaches us that they are classically hidden behind black hole horizons. Piercing through the horizon in quantum gravity is very difficult to do in a controlled environment, with only a handful of (important!) examples \cite{Penington:2019npb,Almheiri:2019psf,Hartman:2013qma,Brown:2015bva}. Since the singularity here is not cloaked behind a horizon, we have the opportunity to learn more on how the quantum dilaton gravity model at hand deals with it. This work can be viewed as a first characterization of this singularity.

We will characterize the near-boundary behavior of sinh dilaton gravity in terms of non-commutative geometry based on the quantum hyperbolic disk, which is a $q$-deformation of the ordinary hyperbolic disk. To that end, we will in this work first investigate precisely what we would expect from a $q$-deformation of the holographic dictionary by solving the $q$-Ward identities, building upon earlier work \cite{Bernard:1989jq,Matsuzaki:1991vu,Almheiri:2024ayc} and extending to our range of $q$. Afterwards, we return to the specific dilaton gravity model at hand and match it to our general predictions. Investigating $q$-holography is an important goal in its own right.

For the related sine dilaton gravity model, a proposal was made for the holographic screen to be in a complexified direction of the geometry \cite{Blommaert:2024ymv}, matching with where the screen is in the analytically continued sinh dilaton gravity model. In this case, this is where the SYK fermions live, and we have a precise link with the underlying microscopic system. As it is good practice to isolate the different issues and treat the difficulties one at a time, we study in this work the sinh dilaton gravity case (which is simpler), leaving a more detailed investigation for the sine dilaton gravity and DSSYK case to future work.

The remainder of this paper is organized as follows.
In \textbf{section \ref{s:qward}}, we review and generalize the bootstrap argument of finding the $q$-CFT two-point function from global conformal $q$-invariance. \textbf{Section \ref{s:bulkq}} defines the bulk gravitational model we will study here. We take the probe regime of the boundary two-point function and match it precisely onto the bootstrapped formula of section \ref{s:qward}. We comment on various routes for the future in the concluding \textbf{section \ref{s:concl}}.
The appendices contain various supplementary material. In particular, \textbf{Appendix \ref{app:dil}} contains an in-depth discussion on 2d dilaton gravity models, the holographic counterterms, and the quantization of sinh dilaton gravity. \textbf{Appendices \ref{app:cft} and \ref{app:form}} contain technical details on the relevant representations and on some mathematical manipulations with $q$-functions.

\section{Correlation functions on the quantum disk}
\label{s:qward}
We first remind the reader of the conventional CFT story. Correlation functions of primary operators in a conformal field theory are strongly constrained by the Ward identities of the corresponding global conformal subgroup. For a 2d CFT, holomorphic (quasi)primary operators with weight $\Delta$ transform under a global conformal PSL$(2)$ Möbius transformation \begin{equation}
z\rightarrow \frac{az+c}{bz+d},
\end{equation}
as
\begin{align}
\label{eq:quasiprimarytransf}
    \mathcal{O}(z)\rightarrow \left(bz+d\right)^{-2\Delta} \;\mathcal{O}\left(\frac{az+c}{bz+d}\right),
\end{align}
where $\Delta$ is the conformal scaling dimension of the quasi-primary operator. The corresponding infinitesimal transformations generate an $\mathfrak{sl}_2$ algebra $[L_0,L_{\pm1}]=\pm L_{\pm1}, \;[L_{+1},L_{-1}]=-2L_0$ with 
\begin{align}
\label{eq:infgenerators}
    L_{-1}=\partial_z, \qquad L_{0}=z\partial_z+\Delta , \qquad L_{+1}=z^2\partial_z+2\Delta z.
\end{align} 
The trivial \emph{coproduct} $\Delta(L_n)=L_n\otimes \mathbf{1}+\mathbf{1}\otimes L_n$ for $n=-1,0,1$ distributes the action of the differential generators on the different operators in a correlation function, leading to the three Ward identities, which demand the correlation functions to remain invariant under the action by the global conformal subgroup:
\begin{align}
\label{eq:distribute}
     L_n\braket{\mathcal{O}(z_1)\mathcal{O}(z_2)}=m\circ\braket{\Delta(L_n)[\mathcal{O}(z_1)\otimes\mathcal{O}(z_2)]}\equiv 0, 
\end{align}
where $m$ is a linear multiplication map which multiplies the entries of the tensor product. Solving these three equations consistently leads to the correlation functions of gravitational theories for which the holographic boundary theory is a standard CFT. 

\subsection{$q$-Correlation functions}

To construct a $q$-deformed version of holography, we analogously imagine the correlators of the bulk gravitational theory to be invariant under the SL$_q(2)$ global subgroup of the dual $q$-deformed conformal boundary theory. To set up the $q$-deformed SL$_q(2)$ conformal subgroup, we claim that the natural generalization of the quasi-primary transformation law \eqref{eq:quasiprimarytransf} acts as a \emph{corepresentation} in a tensor product structure 
\begin{align}\label{eq:corep}
    \mathcal{O}(z) \mapsto  (1\otimes N^{\Delta/2}) \left( z\otimes z_{12} + 1\otimes z_{22} \right)^{-2\Delta} \;\mathcal{O}\left( \frac{z\otimes z_{11} + 1\otimes z_{21}}{ z\otimes z_{12} +1\otimes z_{22}}\right) (1\otimes N^{\Delta/2}).
\end{align}
The coordinates $z_{ij}$ are suitably defined coordinates on the \emph{non-commutative} GL$_q(2)$ manifold for which the determinant $N$ projects down onto the definition of SL$_q(2)$. The meaning of this (co)transformation law is deferred to appendix \ref{app:cft} and is not important for the rest of the main text. Important for us are the analogues of the ``infinitesimal'' generators \eqref{eq:infgenerators}, which can be derived from the above cotransformation rule (see again appendix \ref{app:cft}, where we now redefine $K\rightarrow q^{2L_0}$, $E^+\rightarrow L_{+1}$, $E^-\rightarrow L_{-1}$). They act in terms of scaling operators  $R_qf(z)\equiv f(qz)$ on the holomorphic coordinates, and are given by 
\begin{align}
\label{eq:qprincipalgenerators2}
     L_0=\Delta +z\partial_z, \qquad L_{-1}=\frac{q^{-1/2+\Delta}}{z}\frac{R_{q^2}-1}{q-q^{-1}}, \qquad L_{+1}=-q^{-1/2+\Delta} z\frac{q^{-4\Delta} R_{q^{-2}}-1}{q-q^{-1}},
\end{align} 
which form a realization of the $U_q(\mathfrak{sl}_2)$ algebra, defined by the commutation relations 
\begin{align}
\label{eq:qalgebra}
    [L_0,L_{\pm1}]=\pm L_{\pm1}, \qquad [L_{+1},L_{-1}]=-\frac{q^{2L_0}-q^{-2L_0}}{q-q^{-1}}.
\end{align}
Note that in terms of $q^{2L_0}$, the first relations are also equivalent to $q^{2L_0}L_{\pm1}=q^{\pm2}L_{\pm1}q^{2L_0}$. In the limit $q\rightarrow1$, both the infinitesimal generators and the algebra relations reduce to their classical $U(\mathfrak{sl}_2)$ counterparts.

In a $q$-deformed setup, we again demand the correlation functions to be invariant under the global SL$_q(2)$ conformal group, whose infinitesimal action is generated by the generators \eqref{eq:qprincipalgenerators2}. To distribute the action of the generators on the primary fields in the correlation function as in \eqref{eq:distribute}, we need a more general coproduct which is compatible with the above commutation relations \eqref{eq:qalgebra}.\footnote{When we say a ``compatible'' coproduct, we mean that at the level of the commutation relations, we have 
\begin{align}
\Delta(q^{2L_0}L_{\pm1})=q^{\pm2}\Delta(L_{\pm1}q^{2L_0}),\qquad [\Delta(L_{+1}),\Delta(L_{-1})]=\Delta\left(-\frac{q^{2L_0}-q^{-2L_0}}{q-q^{-1}}\right),
\end{align} 
where by definition of the coproduct, we have $\Delta(AB)=\Delta(A)\Delta(B)$.} A consistent set of coproducts is given by 
\begin{align}
    \Delta(q^{2L_0}) = q^{2L_0} \otimes q^{2L_0}, &\quad \Delta(L_{+1}) =1 \otimes L_{+1} + L_{+1} \otimes q^{-2L_0}, \quad \Delta(L_{-1}) =  q^{2L_0} \otimes L_{-1} + L_{-1} \otimes 1.
\end{align} 
Focusing specifically on two-point correlation functions, leads to the three Ward identities 
\begin{align}
\label{eq:qward}
    m\circ\braket{(\Delta(q^{2L_0})-1)[\mathcal{O}(z_1)\otimes\mathcal{O}(z_2)]}\equiv 0, \quad m\circ\braket{\Delta(L_{\pm1})[\mathcal{O}(z_1)\otimes\mathcal{O}(z_2)]}\equiv 0.
\end{align}  
The explicit recursion relations are then respectively 
\begin{align}
    &0=\braket{\mathcal{O}(z_1)\mathcal{O}(z_2)}-q^{4\Delta}\braket{\mathcal{O}(q^2z_1)\mathcal{O}(q^{2}z_2)},\\
    &0=z_1\left[q^{-6\Delta}\braket{\mathcal{O}(q^{-2}z_1)\mathcal{O}(q^{-2}z_2)}-q^{-2\Delta}\braket{\mathcal{O}(z_1)\mathcal{O}(q^{-2}z_2)}\right] \nonumber \\&\qquad \qquad+z_2\left[q^{-4\Delta}\braket{\mathcal{O}(z_1)\mathcal{O}(q^{-2}z_2)}-\braket{\mathcal{O}(z_1)\mathcal{O}(z_2)}\right], \\&0=\frac{1}{z_1}\left[\braket{\mathcal{O}(q^2z_1)\mathcal{O}(z_2)}-\braket{\mathcal{O}(z_1)\mathcal{O}(z_2)}\right] \nonumber \\&\qquad \qquad+\frac{q^{2\Delta}}{z_2}\left[\braket{\mathcal{O}(q^2z_1)\mathcal{O}(q^{2}z_2)}-\braket{\mathcal{O}(q^2z_1)\mathcal{O}(z_2)}\right].
\end{align}
The effect of a non-trivial coproduct breaks the cyclic permutation symmetry of the correlator, since the distributive action of the generators acts differently on the two coordinates. When both operators share the same conformal weight $\Delta$, a  consistent solution to all three equations is given by a ratio of double sine functions 
\begin{align}
\label{eq:exactcorrelation}
    \boxed{\braket{\mathcal{O}(z_1)\mathcal{O}(z_2)}=\frac{1}{(z_1z_2)^\Delta}\frac{S_b\left(-b\Delta+\frac{\ln(z_1/z_2)}{2\pi ib}\right)}{S_b\left(b\Delta+\frac{\ln(z_1/z_2)}{2\pi ib}\right)}},
\end{align}
when $q=e^{i\pi b^2}$ with $b\in \mathbb{R}$. Indeed, we see that the first recursion relation is trivially satisfied. Using the recursion relation $S_b(x+b)=2\sin(\pi bx) S_b(x)$, the second equation is equivalent to 
\begin{align}
&z_1\left[q^{-2\Delta}\sin\pi b\left(b\Delta+\frac{\ln(z_1/z_2)}{2\pi ib}\right)-\sin\pi b\left(-b\Delta+\frac{\ln(z_1/z_2)}{2\pi ib}\right)\right]\\&+z_2\left[q^{-2\Delta}\sin\pi b\left(-b\Delta+\frac{\ln(z_1/z_2)}{2\pi ib}\right)-\sin\pi b\left(b\Delta+\frac{\ln(z_1/z_2)}{2\pi ib}\right)\right] \overset{?}{=} 0.
\end{align}
Using $q=e^{\pi ib^2}$ this is indeed true. One analogously solves the third equation.

Using the reflection property $S_b(Q-x)=1/S_b(x)$ and the recursive properties of the double sine functions $S_b(x+b)=\sin(\pi bx)S_b(x)$, $S_b(x+1/b)=\sin\frac{\pi x}{b}S_b(x)$, we can rewrite \eqref{eq:exactcorrelation} as:
\begin{align}
    \sim \frac{1}{(z_1z_2)^\Delta}\sin\left(b^2\pi\Delta-i\frac{\ln(z_1/z_2)}{2 }\right)\sin\left(-\pi \Delta +i\frac{\ln(z_1/z_2)}{2 b^2}\right)S_b\left(-b\Delta\pm i\frac{\ln(z_1/z_2)}{2\pi b}\right),
\end{align}
which is a useful form for taking the $b\to 0$ limit. Indeed, upon using the asymptotics $\lim_{b\rightarrow0}S_b(bx\pm i\alpha/\pi b)\sim \sinh(\alpha)^{2x-1}e^{-\alpha/b^2}$, the correlation function limits to the classical conformally invariant two-point function
\begin{align}
    \braket{\mathcal{O}(z_1)\mathcal{O}(z_2)}\sim \frac{1}{(z_1-z_2)^{2\Delta}}.
\end{align}

To obtain a finite temperature solution, we set the boundary coordinates $z_1=e^{i\varphi},\;z_2=e^{i\theta}$. Under the classical conformal transformation, the $q$-independent prefactors vanish, leading to 
\begin{align}
\label{eq:generalsolution}
    \braket{\mathcal{O}(\varphi)\mathcal{O}(\theta)}=\frac{S_b\left(-b\Delta+\frac{\varphi-\theta}{2\pi b}\right)}{S_b\left(b\Delta+\frac{\varphi-\theta}{2\pi b}\right)}=\frac{S_b\left(-b\Delta+\frac{\tau}{\beta b}\right)}{S_b\left(b\Delta+\frac{\tau}{\beta b}\right)},
\end{align}
where the difference in the angular coordinates is related to the temperature $\beta$ as $\varphi-\theta=\frac{2\pi}{\beta}\tau$. Note that the above expressions are not symmetric under $\tau\rightarrow \beta-\tau$. We can make a substitution to ``fake'' coordinates 
\begin{align}
    \varphi\rightarrow\varphi+\frac{\pi b^2}{2}, \qquad \theta\rightarrow\theta-\frac{\pi b^2}{2},
\end{align} 
to obtain the correlator on a ``fake'' quantum disk 
\begin{align}
\label{res}
\braket{\mathcal{O}(\varphi)\mathcal{O}(\theta)}_{\text{fake}}= \frac{S_b\left(-b\Delta+\frac{b}{2}+\frac{\tau}{\beta b}\right)}{S_b\left(b\Delta+\frac{b}{2}+\frac{\tau}{\beta b}\right)}.
\end{align}
Using the recursive property $S_b(Q-x)=1/S_b(x)$, it is readily seen that the expression \eqref{res} now \emph{does} exhibit the KMS symmetry $\tau\rightarrow \beta-\tau$. This is the expression which will be compared below to the probe approximation of the sinh dilaton gravity two-point function \eqref{eq:final}, which inherently preserves the KMS symmetry of the exact answer.

We can also rewrite the exact expression \eqref{eq:generalsolution} suggestively by assuming $\Delta\in \mathbb{N}$. Again using the recursion property $S_b(x+b)=2\sin(\pi bx)S_b(x) $ successively, we arrive at an integer product of shifted sine functions
\begin{align}
\label{2.46}
    \braket{\mathcal{O}(\varphi)\mathcal{O}(\theta)}=\frac{1}{\prod_{n=-\Delta}^{\Delta-1}2\sin\left(\pi b^2n+\frac{\pi\tau}{\beta }\right)}.
\end{align} 
For integer $\Delta$ this finite product can, up to a $\tau$-independent prefactor, also be expressed as 
\begin{align}
\label{eq:qpoch}
    \braket{\mathcal{O}(\varphi)\mathcal{O}(\theta)}\simeq \frac{1}{\left(q^{-2\Delta}e^{\frac{2\pi i}{\beta}\tau};q^2\right)_\Delta\left(q^{-2\Delta+2}e^{-\frac{2\pi i}{\beta}\tau};q^2\right)_\Delta},
\end{align}
where the integer $q$-Pochhammer symbol is defined as $(a,q)_\Delta=\prod_{n=0}^{\Delta-1}(1-aq^n)$. The solution to the Ward identities is then unique depending on the range of $q$. For $q=e^{i\pi b^2}$ on the unit circle, \eqref{eq:generalsolution} is the natural analytic continuation to non-integer $\Delta$. For $0<q<1$, one can analytically continue the $q$-Pochhammer symbols in \eqref{eq:qpoch} to a convergent infinite product as 
\begin{align}\label{eq:analyplch}
     \braket{\mathcal{O}(\varphi)\mathcal{O}(\theta)}=\frac{\left(e^{\frac{2\pi i}{\beta}\tau};q^2\right)_\infty}{\left(q^{-2\Delta}e^{\frac{2\pi i}{\beta}\tau};q^2\right)_\infty}\frac{\left(q^2e^{-\frac{2\pi i}{\beta}\tau};q^2\right)_\infty}{\left(q^{-2\Delta+2}e^{-\frac{2\pi i}{\beta}\tau};q^2\right)_\infty}.
\end{align} 
This solution solves the same $q$-Ward identities \eqref{eq:qward} in this range of $0<q<1$ after going back to real-time coordinates, and matches the old results found by \cite{Bernard:1989jq,Matsuzaki:1991vu}, and more recently revisited in \cite{Almheiri:2024ayc}. This change in structure, depending on whether $q$ is on the unit circle or not, is of a similar spirit as the motivation behind Faddeev's quantum dilogarithm \cite{Faddeev:1999fe}.

In order to find a bulk description of the solution \eqref{res} of the $q$-Ward identities, we consider its spectral decomposition. To this end, we make use of a modified version of Ramanujan's summation formula \eqref{master}. We report it here:
\begin{align}
\label{ram}
\int_\mathbb{R} \mathrm{d}\tau e^{-2\pi \beta \tau} e^{\pi (Q-\alpha)\tau} \frac{S_b(\alpha/2 \pm i \tau)}{S_b(\alpha)}= \frac{S_b(\beta)}{S_b(\alpha+\beta)}.
\end{align}
We can use \eqref{ram} to rewrite our result \eqref{res} in Fourier space as ($\beta_M=b \Delta$):
\begin{align}
\label{eq:ramanexplicit}
    \int_{-\infty}^{\infty}\mathrm{d}\omega\; e^{\frac{\omega \beta}{2b^2}-\frac{\tau\omega}{b^2}}\frac{S_b\left(\beta_M\pm i \frac{\omega\beta}{2\pi b}\right)}{S_b(2\beta_M)}= \frac{S_b(-\beta_M+\frac{b}{2}+\frac{\tau}{b\beta})}{S_b(\beta_M+\frac{b}{2}+\frac{\tau}{b\beta})}.
\end{align}
The LHS of \eqref{eq:ramanexplicit} contains dynamical information about the dual bulk gravitational system as follows. After Fourier transforming the boundary two-point function, one can interpret the poles at $\Im \omega < 0$ in the frequency domain as the quasi-normal modes of the bulk system. This generically evaluates here to just the poles of the double sine function as a double array:\footnote{For $\beta_M \in b \mathbb{N}+\frac{\mathbb{N}}{b}$ there is a partial cancellation of poles with zeros.}
\begin{equation}
\label{qnm}
\omega_{n,m} = -i \frac{2\pi b}{\beta}\left(\beta_M+nb+\frac{m}{b}\right), \quad n,m \geq 0.
\end{equation}
These quasi-normal mode frequencies contribute to the retarded Green's function, and their appearance directly points to a dissipative and thermalizing bulk system, pointing at a black hole horizon. In the AdS$_2$ limit where $b\to 0$, the second set of poles at nonzero $m$ shifts to infinity and is effectively lost. The surviving set of poles just become the known AdS$_2$ black hole quasi-normal mode frequencies (see e.g. \cite{Keeler:2014hba}).

\subsection{Quantum hyperbolic disk}
We have up to this point only discussed the boundary quantum algebra action on boundary correlation functions. Here we show how this almost uniquely fixes the non-commutative geometrical features of the induced bulk region if this is to be the boundary theory. This roughly reverses the arguments of \cite{Berkooz:2022mfk,Almheiri:2024ayc}. To match with conventional notation for quantum groups, we change the notation from the previous subsection into:
\begin{equation}
q^{2L_0} \to K, \qquad  L_{+1} \to E^+, \qquad  L_{-1} \to E^-.
\end{equation}

We introduce a new antiholomorphic coordinate $z^*$, and if we demand that the boundary is at the locus $zz^*=1$, we can deduce the full bulk structure as follows. Firstly, from compatibility of the coproduct 
\begin{gather}
    \Delta(K) = K \otimes K, \quad \Delta(E^+) = \mathbf{1} \otimes E^+ + E^+ \otimes K^{-1}, \quad
    \Delta(E^-) =  K \otimes E^- + E^- \otimes \mathbf{1},
\end{gather}  
with this boundary condition, we can obtain the action of the generators on both bulk coordinates $z$ and $z^*$. The specific generators which generate quantum Möbius transformations on the coordinate $z$ can be found by looking at the spin-0 corepresentation  \eqref{eq:corep}. The associated generators are then given by the induction procedure in Appendix \ref{app:cft}, which act on holomorphic monomials as
\begin{align}
K f(z) &= f(q^2z), \\
E^- f(z) &= \frac{q^{-1/2}}{z}\frac{f(q^{2}z)-f(z)}{q-q^{-1}}, \\
E^+ f(z) &= -q^{-1/2}z \frac{f(q^{-2}z)-f(z)}{q-q^{-1}},
\end{align}
We now treat $z^*$ on the same footing as $z$, and apply these transformations on the boundary constraint $zz^*=1$ directly. This yields e.g.
\begin{equation}
K(z \otimes z^*) \equiv K(z)K(z^*)=1,
\end{equation}
where we used the grouplike coproduct structure $\Delta(K)=K\otimes K$ and $K(c) = 1$ for constants. This fixes then $K (z^*)$ which by holomorphicity holds everywhere in the bulk. This allows us to bootstrap the action of $K,E^+$ and $E^-$ on arbitrary monomials $f(z^*)$.\footnote{We find $K(z^*)=q^{-2}z^*$, $E^+(z^*)=-q^{1/2}$ and $E^-(z^*)=-q^{-3/2}{z^*}^2$. By successively using the coproduct to find the action on $z^{*^n}$, one deduces (\ref{eq:staractions}) - \eqref{eq:staractions2}.  } We obtain
\begin{align}
\label{eq:staractions}
K f(z^*) &= f(q^{-2} z^*), \\
E^- f(z^*) &= q^{-1/2}z^* \frac{f(q^{-2}z^*)-f(z^*)}{q-q^{-1}}, \\
\label{eq:staractions2}
E^+ f(z^*) &= -\frac{q^{-1/2}}{z^*} \frac{f(q^{2}z^*)-f(z^*)}{q-q^{-1}}.
\end{align}
These three relations, together with the previous ones acting on $z$, implement the action of the quantum group transformations on the two bulk coordinates $z$ and $z^*$. Secondly, ansatzing a non-commutativity relation $z^*z = Azz^* + B$ for $c$-numbers $A$ and $B$, and working out e.g. $E^-(z^*z) = E^-(Azz^*+ B)$ via the coproduct immediately fixes $A$ and $B$ as:
\begin{equation}
\label{eq:noncom}
 z^*z = q^2 zz^* + 1-q^2.
\end{equation}
This means that $z^*z=1=zz^*$ describes the boundary, where hence these two coordinates commute. To sum up, the boundary U$_q(\mathfrak{sl}(2,\mathbb{R}))$ quantum algebra symmetry, implies the bulk \emph{needs} to have non-commutative coordinates \eqref{eq:noncom} to be compatible.

\begin{figure}[h]
    \centering
    \includegraphics[width=0.27\linewidth]{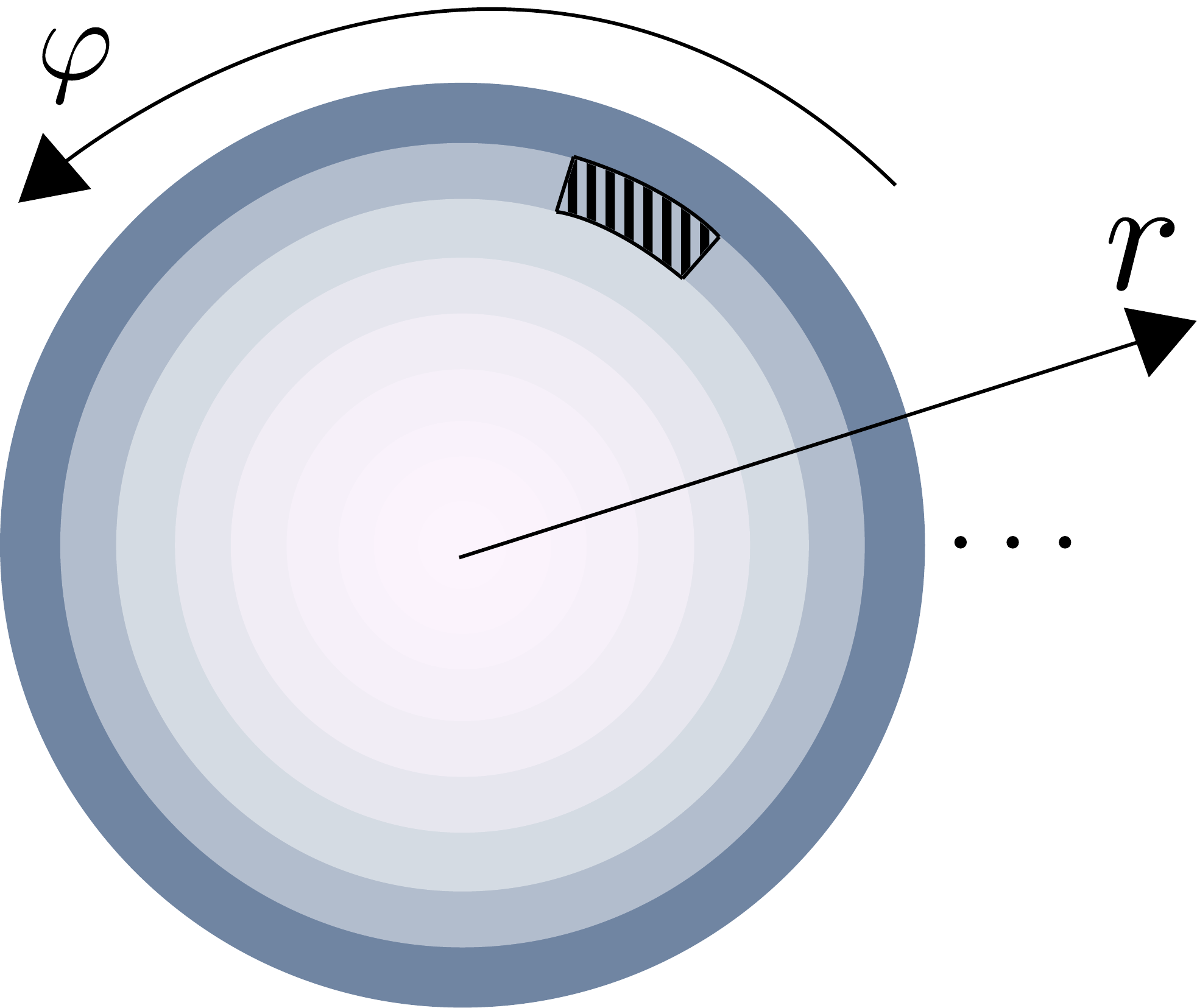}
    \caption{Quantum hyperbolic disk with fuzzy cells of size $2\abs{\log q}$ in $(r,\varphi)$ coordinates where $r \in [0,+\infty[$ and $\varphi\in [0,2\pi[$ where the cell sizes are uniform.}
    \label{FigQHD}
\end{figure}

These two coordinates can be conveniently written by introducing a radial coordinate $y=1-zz^*$ \cite{Almheiri:2024ayc}, with non-commutativity:
\begin{equation}
zy=q^{-2}yz, \qquad z^*y=q^{2}yz^*.
\end{equation}
The $y$-coordinate classically ranges between 1 (the origin), and 0 (the boundary). Finally introducing new coordinates as $y=e^{-r}$ (where $r$ classically ranges from 0 (the origin) and $+\infty$ (the boundary), and $z^{*^{-1}}z=e^{2i\varphi}$, we obtain
\begin{equation}
[\varphi,r] = -2i \log q .
\end{equation}
If we would have taken $0<q<1$, this would be a standard noncommutativity relation, telling us that spacetime is fundamentally fuzzy with elementary (Planckian) cell size of $2 \abs{\log q}$ (see Fig.~\ref{FigQHD}).\footnote{The coordinate $\varphi$ is periodic with period $2\pi$, which implies the conjugate variable $r/2\abs{\log q}$ is discretized $\in \mathbb{N}$. This is identical to the DSSYK discrete chord number.} 
In the following, we will instead have $q$ on the unit circle, and the above fuzzyness is still there, but furthermore complexified.

Just as JT gravity provides an explicit realization of $\mathrm{AdS}_2$ spacetime within a quantum gravity model, can we similarly embed the quantum hyperbolic disk into a fully dynamical bulk gravitational system?

\section{A candidate bulk description for $q$-holography }
\label{s:bulkq}
In the previous section, we derived the $q$-conformal two-point function by constraining its form \eqref{res} through the requirement of invariance under the global part of the $q$-conformal group. Furthermore, we interpreted its Fourier transform as encoding information about the quasi-normal modes \eqref{qnm} of a dual bulk system. This observation hints at the presence of a black hole in the bulk, whose properties we now seek to identify. 

\subsection{Definition and classical structure of sinh dilaton gravity}
\label{s:classical}
The most natural approach is to look for a bulk gravitational theory whose asymptotic symmetries reproduce the SL$_q(2)$ algebra, which was used to constrain the two-point function. As we shall see, it turns out that the most natural candidate is Liouville gravity \cite{Polyakov:1981rd,Distler:1988jt,David:1988hj,Knizhnik:1988ak}. This is a string theory consisting of a spacelike Liouville CFT, a matter CFT and the usual bc ghost system ensuring cancellation of the conformal anomaly. If we select a timelike Liouville CFT as the matter sector, then we have a simple Lagrangian description. Through a field redefinition \cite{Kyono:2017pxs,Mertens:2020hbs,Fan:2021bwt}, this theory can be reformulated as a two-dimensional dilaton gravity model with a $\sinh$ profile for the dilaton potential. Concretely, the theory on a manifold $\mathcal{M}$ with holographic boundary $\partial \mathcal{M}$ is defined by the Euclidean action:
\begin{equation}\label{Liouville}
S = -\frac{1}{2} \int_\mathcal{M} d^2x \sqrt{g}\left(\Phi R + \frac{\sinh 2\pi b^2 \Phi}{\pi b^2}\right) - \oint_{\partial \mathcal{M}} d \tau\,\sqrt{h}\,\bigg(\Phi K-\sqrt{\frac{\cosh(2\pi b^2 \Phi)-1}{2\pi^2 b^4}}\bigg).
\end{equation}
The boundary term contains both the GHY boundary term, and a holographic counterterm required to obtain a finite on-shell action that is essentially unique. We provide a detailed discussion on this counterterm in general dilaton gravity models in Appendix \ref{a:counter}.

Relatedly, in \cite{Collier:2023cyw} the above sinh dilaton gravity model was defined and studied within the framework of topological recursion and random matrix theory,\footnote{When introducing a boundary, they use a different boundary term than us, and hence their amplitudes (e.g. the disk partition function) would differ from ours.} of direct relevance to understand wormhole and multiboundary correlation functions. We will stick with the lowest disk topology in this work.

As a consistency check, we note that in the $b \rightarrow 0$ limit, the full action in \eqref{Liouville} reduces to the JT action $+$ boundary term. The deformation parameter $b$, which controls the deviation from JT gravity, is related to the central charge of the spacelike Liouville field $\varphi$ in Liouville gravity via $c_{\varphi} = 1 + 6Q^2$, where $Q = b + 1/b$, and to the quantum group deformation parameter as $q=e^{i \pi b^2}$.

Using the Poisson sigma model reformulation of $\sinh$ dilaton gravity \cite{Fan:2021bwt}, one can indeed show that the theory \eqref{Liouville} in the first-order formalism induces the desired SL$_q(2)$ algebra on the boundary \cite{Blommaert:2023wad}. One can alternatively stay in the second-order formulation and quantize the model directly from there. We collect and complete this approach in Appendix \ref{a:quant}. This then leads to exact expression for the disk partition function and the boundary correlation functions.

Before analyzing the boundary two-point function of interest in this work, it is instructive to examine the classical geometry of sinh-dilaton gravity and its connection to the thermodynamics of the corresponding black hole. The classical metric and dilaton solution in sinh dilaton gravity \eqref{Liouville} is given by
\begin{equation}
\label{geom}
\mathrm{d} s^2 = - \frac{1}{2\pi^2b^4} \left[\cosh 2\pi b^2 r - \cosh 2 \pi b^2 r_h \right] dt^2 + \frac{2\pi^2b^4 dr^2}{ \left[\cosh  2\pi b^2 r - \cosh 2\pi b^2 r_h \right]}, \quad \Phi(r) = r.
\end{equation}
This spacetime describes a geometry with a black hole horizon at $r = r_h \equiv \Phi_h$, and we take the holographic boundary to be located at $r = +\infty$, where the metric diverges. Varying the action with respect to the boundary metric yields the following expression for the ADM energy of the theory:
\begin{equation}
\label{adm}
E_{\mathrm{ADM}} = \frac{\cosh (2\pi b^2 \Phi_h)}{4\pi^2 b^4},
\end{equation}
where a divergent contribution has been canceled by the holographic counterterm included in \eqref{Liouville}. The associated Hawking temperature is 
\begin{equation}
\label{temperature}
\beta_{\mathrm{BH}}=4\pi^2b^2 /\sinh(2\pi b^2 \Phi_h).
\end{equation}

In the deep interior where $r,r_h \ll b^{-2}$, we can expand the $\cosh$ functions in \eqref{geom} and we recover the black hole patch of the JT gravity metric solutions.
Far away at large $r$, the geometry is very different, and we reach a curvature singularity at $r\to +\infty$, since the scalar curvature is given by
\begin{equation}\label{R}
R = - 2\cosh 2\pi b^2 r.
\end{equation}
As noted in the introduction, this is the same type of singularity encountered in sine-dilaton gravity when approaching the holographic screen at $\Phi = \pi/2 + i\infty$, where the DSSYK hologram resides \cite{Blommaert:2024ymv,Blommaert:2024whf}.

Within our conventions and notation, the exact disk partition function of sinh dilaton gravity is given by \cite{Mertens:2020hbs}:
\begin{align}
\label{partfuncs}
Z(\beta) &=~ \begin{tikzpicture}[baseline={([yshift=-.5ex]current bounding box.center)}, scale=0.5]
\draw[fill=blue!40!white,opacity=0.3] (0,0) ellipse (1.5 and 1.5);
\node at (-2,0) {\small $\beta$};
\end{tikzpicture} = \int_{0}^{\infty} \mathrm{d}s ~\sinh 2\pi b s \sinh\frac{2\pi s}{b} e^{-\beta \frac{\cosh(2\pi b s)}{4\pi^2b^4} }.
\end{align}
Changing coordinates to the known ADM energy \eqref{adm} with $s\equiv b\Phi_h$, we recognize the spectral density as
\begin{equation}
\rho(E) \sim \sinh\left(\frac{1}{b^2}\text{arccosh}( 4\pi^2b^4E)\right).
\end{equation}
From \eqref{partfuncs} we can extract the saddle point equation:
\begin{equation}
\label{firstlaw}
\sqrt{E^2-\frac{1}{(4\pi^2b^4)^2}} = \frac{1}{b^2 \beta},
\end{equation}
which is a good  semi-classical approximation in the $b\to 0$ regime,\footnote{One should make a distinction between physical quantization and $q$-deformed ``quantization''. Reinstating the quantum mechanical $\hbar$ parameter, relating the classical Poisson brackets and the quantum commutation relations, is equivalent to setting $b^2\rightarrow \hbar b^2$. Taking the semiclassical limit $\hbar\rightarrow0$ in units where $\hbar=1$ is equivalent to taking $b\rightarrow0$ in rescaled coordinates of the dilaton, see appendix \ref{a:quant}. } where $b^4E = \cosh(2\pi bs)/4\pi^2$ and $\beta/b^2$ are kept fixed. The saddle-point relation \eqref{firstlaw} can be interpreted as the connection between the microcanonical and canonical ensembles in sinh dilaton gravity. Indeed it matches the first law of black hole thermodynamics for the sinh dilaton gravity black hole, with energy \eqref{adm} and inverse temperature \eqref{temperature}.

For IR energies close to the cutoff $\frac{1}{4\pi^2b^4}$, i.e. $E \approx \frac{1}{4\pi^2b^4} + E_{\text{JT}}$, we recover the JT black hole first law $E_{\text{JT}} = 2\pi^2 \beta^{-2}$ and the JT density of states: $\sinh\left(\frac{1}{b^2}\text{arccosh}( 4\pi^2b^4E)\right)\sim \sinh(2\pi \sqrt{2E_{\text{JT}}})$. However, for high energies, the behavior is very different $E \sim \beta^{-1}$. Physically, the model has a slower power-law $\sim E^{1/b^2}$ high energy increase of the density of states than the JT Cardy growth $\sim e^{2\pi\sqrt{2E}}$.
These observations illustrate that the usual UV/IR link in holography \cite{Susskind:1998dq} is still present in the sinh dilaton gravity model, but also that the UV regime is quite distinct from asymptotically AdS holography. We worked here in dimensionless units. Restoring units, large energy means that $ E \gg \frac{1}{aG_N}$, where $a$ has units of length and is the dilaton slope near the boundary $\Phi \approx a r$. A sketch of the situation is given below in Fig.~\ref{FigUVIR}.
\begin{figure}[h]
    \centering
    \includegraphics[width=0.55\linewidth]{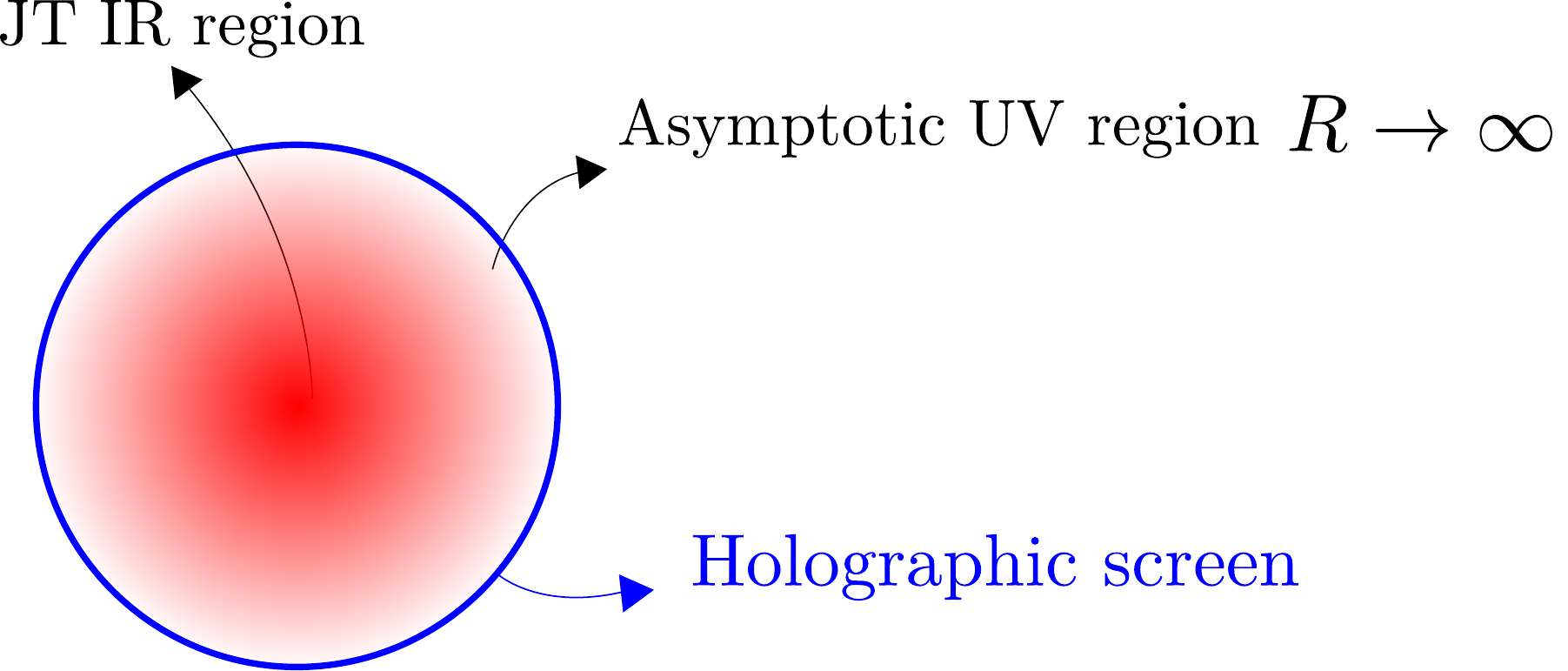}
    \caption{Geometry of the classical sinh dilaton gravity solution, with a JT region deep in the interior, and a UV region close to the holographic screen where the Ricci scalar blows up $R\to-\infty$. The classical solution has an emergent SL$(2,\mathbb{R})$ isometry.}
    \label{FigUVIR}
\end{figure}
Based on the above considerations, taking the high-energy limit, or high-temperature limit because of \eqref{firstlaw}, corresponds to getting closer and closer to the holographic boundary, where it is reasonable to expect that quantum geometry features may resolve the classical curvature singularity \eqref{R} of the sinh dilaton gravity black hole.
Therefore our aim is to probe this UV near boundary regime by injecting a small mass probe into the bulk, and identifying the geometrical features it observes. To that effect, in the next section, we will analyze the sinh dilaton gravity boundary two-point function to investigate whether the structure observed in \eqref{eq:ramanexplicit} emerges in the probe regime. 

To set the stage, we begin with the exact expression for the boundary two-point function \cite{Mertens:2020hbs}:
\begin{align}\label{two-point}
    \begin{tikzpicture}[baseline={([yshift=-.5ex]current bounding box.center)}, scale=0.5]
\draw[fill=blue!40!white,opacity=0.3] (0,0) ellipse (1.5 and 1.5);
\draw[fill] (-1.5,0) circle (0.06);
\draw[fill] (1.5,0) circle (0.06); 
\node at (-2.6,0) {\small $\mathcal{O}_{\beta_M}$};
\node at (2.6,0) {\small $\mathcal{O}_{\beta_M}$};
\node at (0,-1.8) {\small $\tau$};
\node at (0,1.8) {\small $\beta-\tau$};
\end{tikzpicture}
    =
    \int  \mathrm{d} s_1 \mathrm{d} s_2 \rho(s_1) \rho(s_2)\hspace{0.05cm} e^{-E(s_1)\tau} e^{-E(s_2)(\beta-\tau)}\hspace{0.05cm}\frac{S_b\left(\beta_M \pm i s_1 \pm i s_2\right)}{S_b(2\beta_M)},
\end{align}
where $\tau=\left|\tau_1-\tau_2\right| $ is the Euclidean separation between the boundary insertion points of the boundary operators, and $\beta-\tau$ its complement along the thermal circle. Here $E(s) \equiv \frac{\cosh(2\pi b s)}{4\pi^2b^4}$ and $\rho(s) = \sinh(2\pi b s)\, \sinh\left(\frac{2\pi s}{b}\right)$. This expression arises from the canonical quantization of sinh dilaton gravity, as we detail and partially review in appendix \ref{a:quant}, where the spectral density and matrix elements are given in \eqref{spectral} and \eqref{matrix} respectively, and the ADM energy defined in \eqref{adm} appears in the Boltzmann factors (upon identifying $s=b\Phi_h$).
We point out that the same amplitude can be obtained as the boundary tachyon two-point function in the Liouville gravity formulation $ \braket{\mathcal{B}_{\beta_M}\mathcal{B}_{\beta_M}}_{\tau,\beta-\tau}$ with conformal weight of the matter CFT operators $\Delta_{\beta_M}=\beta_M(b-1/b+\beta_M)$, when rewritten in a fixed length basis \cite{Mertens:2020hbs}. 

\begin{figure}[h]
\centering
\begin{tikzpicture}
  \node[anchor=south west,inner sep=0] (image) at (0,0) {\includegraphics[width=0.5\textwidth]{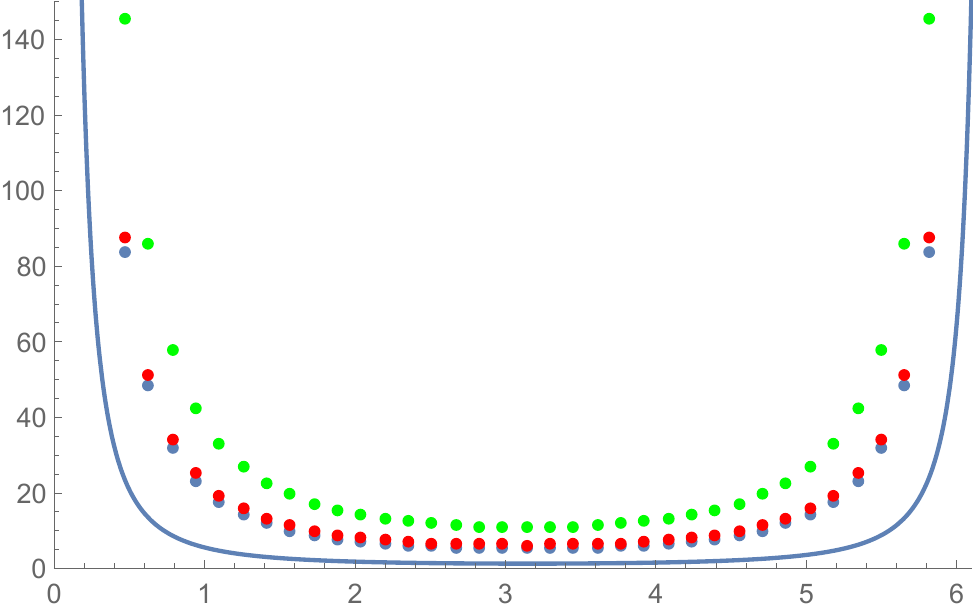}};
  \begin{scope}[x={(image.south east)}, y={(image.north west)}]
    \node at (0.9, -0.05) {\large \textbf{$\tau$}};  
  \end{scope}
\end{tikzpicture}
\caption{Numerical evaluation of the sinh dilaton gravity two-point function $\langle \mathcal{O}_{\beta_M}(\tau)\mathcal{O}_{\beta_M}(0)\rangle_\beta$ \eqref{two-point} as a function of $\tau$, for $\beta=2\pi$ and $\Delta=1$. The green, red, and blue dots correspond to $b = 0.9$, $b = 0.8$, and $b = 0.6$ respectively. The numerical evaluation was obtained using expression \eqref{A.3} following the manipulations described in Appendix \ref{app:form}. The blue line corresponds to the semiclassical ($\hbar \rightarrow 0$) limit \eqref{eq:semicl}.}
\label{fig:numerical}
\end{figure}

The semi-classical limit of this expression \eqref{two-point} was studied in \cite{Blommaert:2023wad}. This corresponds to either taking $\hbar\to 0$ (or equivalently $b \to 0$ upon rescaling the field $\pi b^2 \Phi \to \Phi$). This results in the expression:
\begin{equation}
\label{eq:semicl}
\langle \mathcal{O}_{\beta_M}(\tau)\mathcal{O}_{\beta_M}(0)\rangle_\beta  \quad\overset{\hbar\to 0}{\longrightarrow} \quad \frac{\left(\frac{2 \pi ^2 b^2}{\beta }\right)^{2 \Delta}}{\sin ^{2 \Delta}\left(\frac{\pi  \tau }{\beta }\right)},
\end{equation}
which is just a thermal AdS$_2$ answer. So this limit exhibits explicitly the SL$(2,\mathbb{R})$ isometries of an AdS$_2$ hyperbolic disk, but where is this AdS$_2$ geometry hidden? It is actually a rescaled classical metric
\begin{equation}
\label{eq:weylliou}
\mathrm{d}s_{\mathrm{eff}}^2 \equiv \mathrm{d}s^2 \, e^{-2\pi b^2 \Phi},
\end{equation}
which with the above classical metric solution \eqref{geom} can be readily checked to satisfy $R_{\text{eff}} = -2$. This rescaled metric is at the same time also the underlying Liouville metric in the original Liouville gravity formulation of the problem. This emergent SL$(2,\mathbb{R})$ symmetry algebra is a property of the classical solutions, which is broken in the full quantum answer \eqref{two-point}. A numerical evaluation of the two-point function \eqref{two-point} and its semi-classical regime is presented in Fig.~\ref{fig:numerical}.

\subsection{Probe regime of sinh dilaton gravity}
\label{s:probe}

We now turn to a different regime, the probe regime.
To study it, it is beneficial to change coordinates from the momentum labels $s_1,s_2$ to the mass of the background black hole $E$ and the energy of the probe $\omega$ in terms of 
\begin{align}
\label{eq:transformation}
          E \equiv \frac{1}{4\pi^2b^4}\cosh(2\pi bs_2), \qquad E+\omega/b^2 \equiv \frac{1}{4\pi^2b^4} \cosh(2\pi bs_1).
\end{align} 
Such types of approximations were extensively studied in JT gravity in \cite{Lam:2018pvp}. The Jacobian of this transformation is given by 
\begin{equation}\mathrm{d}M \mathrm{d}\omega\simeq \sinh(2\pi bs_1)\sinh(2\pi bs_2)\mathrm{d}s_1 \mathrm{d}s_2,
\end{equation}
such that the boundary two-point function is equivalent to 
\begin{align}
    \begin{tikzpicture}[baseline={([yshift=-.5ex]current bounding box.center)}, scale=0.5]
\draw[fill=blue!40!white,opacity=0.3] (0,0) ellipse (1.5 and 1.5);
\draw[fill] (-1.5,0) circle (0.06);
\draw[fill] (1.5,0) circle (0.06); 
\node at (-2.6,0) {\small $\mathcal{O}_{\beta_M}$};
\node at (2.6,0) {\small $\mathcal{O}_{\beta_M}$};
\node at (0,-1.8) {\small $\tau$};
\node at (0,1.8) {\small $\beta-\tau$};
\end{tikzpicture}
    =\int_\kappa^\infty \mathrm{d}E\int_{(\kappa-E)b^2}^\infty \mathrm{d}\omega \ \mathcal{A}(E,\omega),
\end{align} 
where the spectral decomposition of the amplitude $\mathcal{A}(E,\omega)$ is given by
\begin{align}
\label{eq:start}
\mathcal{A}(E,\omega)=\sinh\left(\frac{2\pi}{b}s_1 (E,\omega)\right)\sinh\left(\frac{2\pi}{b}s_2 (E,\omega)\right)e^{-\beta E-\frac{\tau}{b^2}\omega}\frac{S_b\left(\beta_M\pm i s_1 (E,\omega)\pm i s_2 (E,\omega)\right)}{S_b(2\beta_M)},
\end{align} 
with the variables $s_1 (E,\omega)$ and $s_2 (E,\omega)$ given by the inverses of \eqref{eq:transformation}. The energy cutoff is defined as $\kappa\equiv \frac{1}{4\pi^2b^4}$. 

In the probe approximation, we imagine a large background energy $E$ which remains unchanged by the injection of the probe, which is assumed not to alter the underlying semiclassical geometry. This is indeed the region of the Hilbert space, corresponding to large energies $E_1,E_2 \gg1$ and a small energy difference, i.e. $E_1\simeq E_2$, where the features of $q$-holography are expected to emerge, as we are approaching the holographic screen of sinh dilaton gravity. We now focus on this regime of the Hilbert space directly at the level of $\mathcal{A}(E,\omega)$ to check whether the structure of the integrand on the left-hand side of \eqref{eq:ramanexplicit} is recovered. In the limit of large energies, we can approximate $\mathcal{A}(E,\omega)$  to first order in the probe energy $\omega$, leading to 
\begin{align}
        s_1&=\frac{1}{2\pi b}\text{arccosh}(4\pi^2 b^4E), \\
        s_2&=\frac{1}{2\pi b}\text{arccosh}\left(4\pi^2b^4 (E+\omega/b^2)\right) \approx \frac{1}{2\pi b}\text{arccosh}(4\pi^2b^4 E)+\frac{2\pi b\omega}{\sqrt{(4\pi^2b^4E)^2-1}}+\dots \nonumber
\end{align} 
By using the asymptotics of the double-sine functions  
\begin{equation}
S_b(x) \to e^{\pm i \frac{\pi}{2}x(Q-x)}e^{\pm i \delta(b)}, \qquad \text{Im}(x) \to \pm \infty,
\end{equation}
with $\delta(b)$ an $x$-independent phase, the product of the two $S_b$ functions involving the total momentum can be approximated as 
\begin{align}
    S_b(\beta_M\pm i(s_1+s_2)) \,\, \rightarrow \,\,e^{2\pi \beta_M(s_1+s_2)-\pi Q (s_1+s_2)}.
\end{align}
Implementing these approximations in \eqref{eq:start}, we get 
\begin{align}
\mathcal{A}(E,\omega) \,\,&\simeq \,\, e^{\frac{1}{b^2} \text{arccosh}\left(4\pi^2b^4 E\right)- \beta E+{\color{darkblue}\frac{2\beta_M}{b}\text{arccosh}\left(4\pi^2b^4E\right)-\text{arccosh}\left(4\pi^2b^4E\right)}} \label{eq:Eintegral}\\
&\times \,e^{\frac{2\pi^2\omega}{\sqrt{(4\pi^2b^4E)^2-1}}-\frac{\tau}{b^2}\omega{\color{darkblue}+\frac{4\pi^2\beta_M b\omega}{\sqrt{(4\pi^2b^4E)^2-1}}-\frac{2\pi^2b^2\omega}{\sqrt{(4\pi^2b^4E)^2-1}}}}\frac{S_b\left(\beta_M\pm i \frac{2\pi b\omega}{\sqrt{(4\pi^2b^4E)^2-1}}\right)}{S_b(2\beta_M)} ,\label{eq:omegaintegral}
\end{align}
where we suggestively bundled the factors together in the two lines. We recognize in the first two terms in the exponent of \eqref{eq:Eintegral} the same terms that would appear in the saddle approximation of the disk partition function \eqref{partfuncs}. For the probe approximation to be successful, the matter operator should be light enough not to disturb the saddle of the background geometry \eqref{firstlaw}. Implementing the \emph{light probe} approximation 
\begin{equation}
\frac{2\beta_M}{b} -1 \ll 1/b^2
\end{equation}
on \eqref{eq:Eintegral}, the two last terms (in darkblue) in the exponent of \eqref{eq:Eintegral} can be discarded. In this way the $E$-saddle is still given by 
\begin{align}
\label{saddle2}
    \sqrt{(4\pi^2b^4E)^2-1} = \frac{4\pi^2b^2}{\beta},
\end{align} 
where the first term in the square root dominates in the large $E$ regime. It is important to note that although the probe approximation does not alter the saddle, the saddle is not necessarily dominating the correlator. In other words, we do not consider a semiclassical $b\rightarrow 0$ regime for which the integrand peaks sharply around the saddle point value. Instead, we are probing a quantum mechanical regime with finite $b$ in which the saddle point experiences large quantum fluctuations.
We can then transition to the microcanonical two-point function of the full quantum correlation function by an inverse Laplace transform to remove the integral over background energies and finally set the energy of the microcanonical correlator to be equal to the large saddle point value $E(\beta)$ given by \eqref{saddle2}. 

Within the probe regime, the two darkblue terms in \eqref{eq:omegaintegral} are also subdominant. We end up with the final microcanonical answer (for which we strip off the contributions on the first line \eqref{eq:Eintegral})
\begin{align}
  \boxed{  \mathcal{A}(E,\omega) \approx  e^{\frac{\omega \beta(E)}{2b^2}-\frac{\tau\omega}{b^2}}\frac{S_b\left(\beta_M\pm i \frac{\omega\beta(E)}{2\pi b}\right)}{S_b(2\beta_M)}},
\end{align}
where we encoded the energy dependence within the saddle temperature equation. 

\begin{figure}[h]
    \centering
    \includegraphics[width=0.3\linewidth]{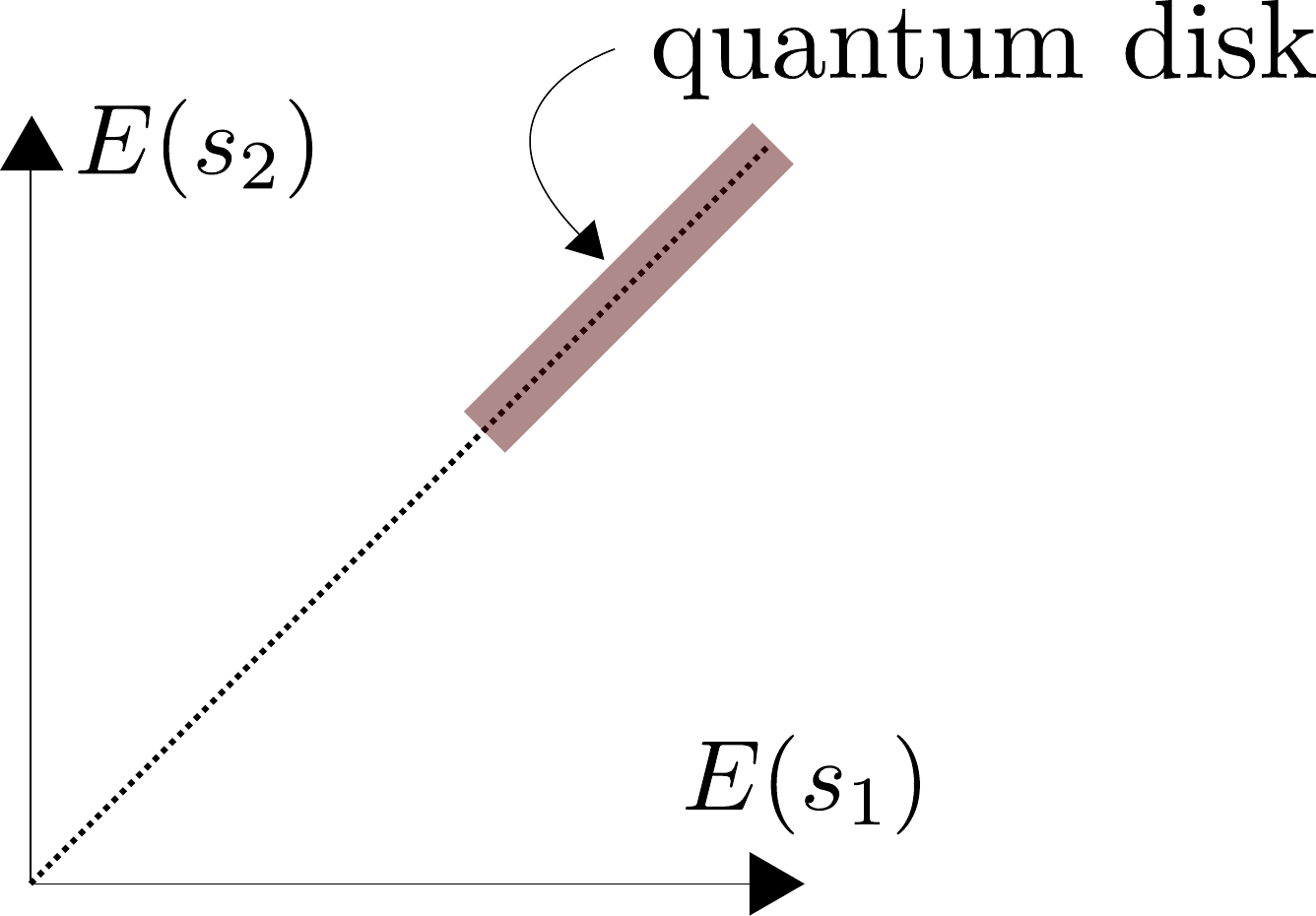}
    \caption{Kinematic regime in frequency space where the probe approximation holds, and where the quantum disk geometry and its isometries govern the correlation functions.}
    \label{Figkinematic}
\end{figure}

The main result of this paper is that this is just the spectral decomposition of the two-point correlator \eqref{eq:ramanexplicit} we derived by studying $q$-Ward identities, but now emerging within this specific kinematic regime of the sinh dilaton gravity two-point function. Hence we explicitly find the quantum hyperbolic disk in the probe region of frequency space (see Fig.~\ref{Figkinematic}).

While the probe integral transform \eqref{eq:ramanexplicit} runs over energies $\omega$ from $-\infty$ to $+\infty$, for the probe approximation to be valid on the bulk side, we require $\omega \ll E$. Even when $E$ is taken to be very large, as in our case, for any fixed $E$ there will inevitably be a region in the Hilbert space where $\omega \sim \mathcal{O}(E)$, at which point the probe approximation breaks down.
We may thus conclude that the quantum disk accurately captures the sector of the sinh dilaton gravity Hilbert space corresponding to the probe regime at large energies. However, a complete identification in coordinate space would require extrapolating the probe regime beyond its domain of validity. Under this extrapolation, we would obtain\footnote{As a technical aside, the particular ratio of $S_b$-functions in \eqref{eq:final} has appeared before in the literature and has nice structural properties. In \cite{Teschner:2012em}, it would be denoted as $D_{-i\beta_M}(i\frac{\tau}{b\beta}+i\frac{1}{2b})$, and was identified as the kernel of the intertwiner $R_\alpha: \mathcal{P}_\alpha \to \mathcal{P}_{Q-\alpha}$, mapping Virasoro representations $\alpha$ into their conjugate $Q-\alpha$. This same function was then identified as the building block of the $3j$-symbols of the modular double \cite{Teschner:2012em}.}
\begin{equation}
\label{eq:final}
\boxed{\left\langle \mathcal{O}(\tau) \mathcal{O}(0)\right\rangle_{\beta(E)} \quad\overset{\text{probe}}{\to} \quad \frac{S_b(-\beta_M+\frac{b}{2}+\frac{\tau}{b\beta(E)})}{S_b(\beta_M+\frac{b}{2}+\frac{\tau}{b\beta(E)})}}.
\end{equation}
We have considered very large background energies with light probes that do not influence the thermodynamical saddle. Due to the UV-IR relation in this model (as discussed above in subsection \ref{s:classical}), this kinematic regime probes the geometric region close to the boundary of the dilaton gravity model.

\begin{figure}[h]
    \centering
    \includegraphics[width=0.45\linewidth]{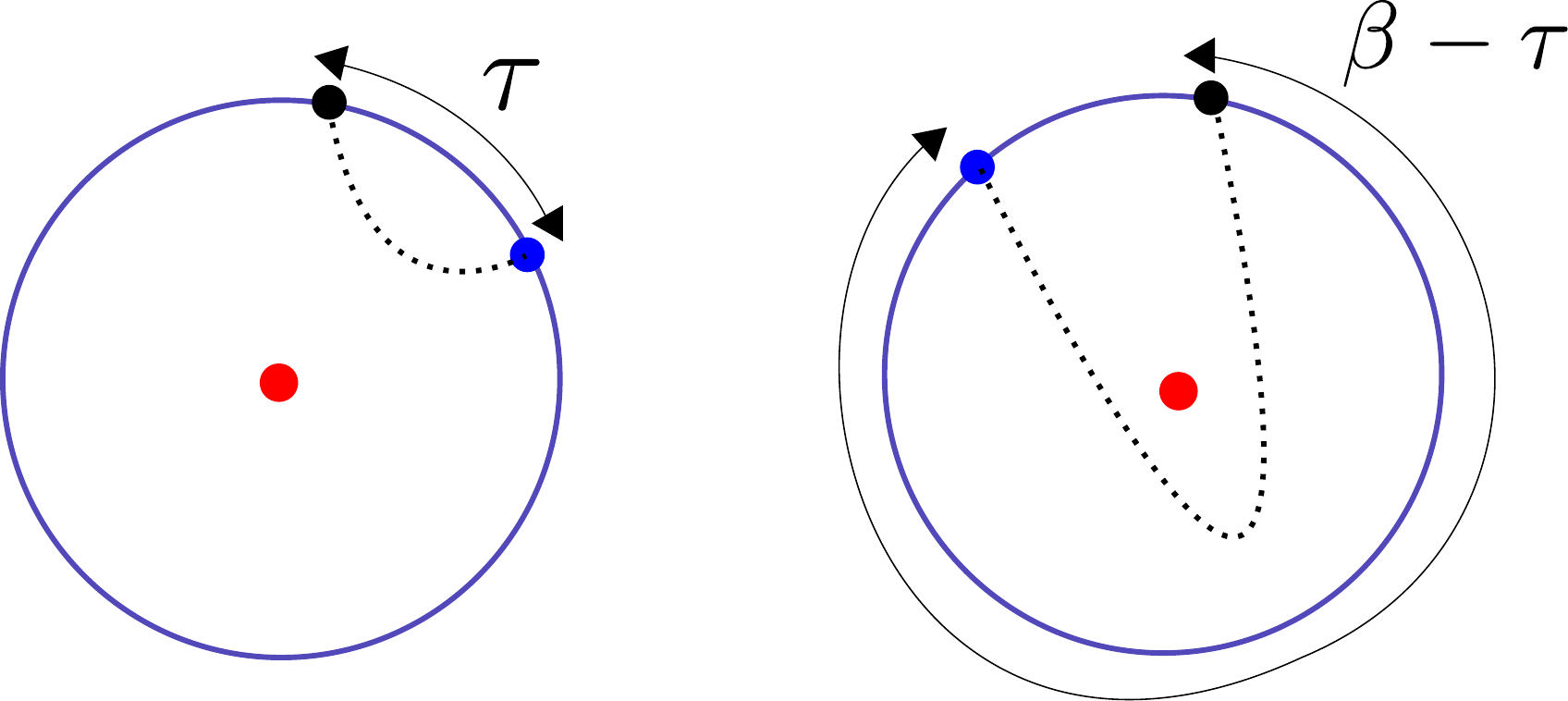}
    \caption{The KMS condition requires that there is no defect in the geometrical bulk. The blue endpoint is moved around the boundary to a distance $\beta-\tau$, but it would pick up any defect in the interior, as illustrated by the dashed line now encircling the red defect (if it would be there).}
    \label{FigKMS}
\end{figure}

If we interpret this result as a saddle evaluation of a canonical correlation function $\text{Tr}( \mathcal{O}(\tau) \mathcal{O}(0)e^{-\beta H})$, using $S_b(Q-x)=1/S_b(x)$, then our expression \eqref{eq:final} manifestly preserves the KMS condition $\tau\rightarrow \beta-\tau$ of the original Liouville correlator, as required for any thermal correlation function. This means physically that there is no defect in the interior of the thermal disk (see Fig.~\ref{FigKMS}).

Assuming $\beta_M\equiv b\Delta$ with $\Delta$ integer, we can use the recursive property of the $S_b(.)$ functions $S_b(x+b)=2\sin(\pi bx)S_b(x)$ to rewrite the correlation function in terms of elementary functions as ($\beta_M = b\Delta$):
 \begin{align}
 \label{eq:sine}
    \frac{S_b(-\beta_M+\frac{b}{2}+\frac{\tau}{b\beta})}{S_b(\beta_M+\frac{b}{2}+\frac{\tau}{b\beta})}=\frac{1}{\prod_{n=-\Delta}^{\Delta-1}2\sin\left(\frac{\pi\tau}{\beta}+\frac{\pi b^2}{2}+\pi b^2n\right) },
\end{align} 
Again taking the JT limit $b\to 0$ readily leads to the usual thermal CFT$_1$ answer:
\begin{equation}\label{eq:classicalcorrelator}
\left\langle \mathcal{O}(\tau) \mathcal{O}(0)\right\rangle_{\beta}\,\,\overset{\text{probe, } b\to 0}{\longrightarrow} \,\, \frac{1}{\left(\sin \frac{\pi}{\beta}\tau\right)^{2\Delta}}.
\end{equation}
The full probe answer \eqref{eq:sine} has $2\Delta$ simple poles at locations 
\begin{equation}\label{poles}
\tau_{\text{pole}} =\beta b^2\Delta, \, \beta b^2(\Delta-1), \, \hdots  , \, 0, \,  \hdots , \, -\beta b^2(\Delta-1).
\end{equation}
and their shifts by multiples of $\beta$. The fact that the composite pole is disentangled into simple poles is intuitively associated to the fuzzyness of non-commutative geometry. However, Fig.~\ref{fig:numerical} shows that the exact sinh dilaton gravity two-point function does not exhibit these extra poles \eqref{poles} in $\tau$. These poles arise as a result of the probe extrapolation. Heuristically, this can be explained by noting that the other poles would correspond to bulk trajectories that extend too deep into the bulk, where the quantum disk approximation is no longer valid.

The fact that the semi-classical answer ($\hbar \to 0$) and the probe answer (UV or near-boundary region) disagree with each other can be physically interpreted in that the near-boundary region is not a classical geometry; it is instead the quantum hyperbolic disk. Where is this quantum hyperbolic disk geometry hidden?
This emergent quantum hyperbolic disk geometry should be viewed as a $q$-deformation of the effective AdS$_2$ geometry that governed semi-classical physics \eqref{eq:weylliou}. To see this, it is useful to rewrite the sinh dilaton gravity model in terms of $q$-Liouville quantum mechanics, as explained in detail in Appendix~\ref{a:quant}. One of the phase space variables is the length $L$ of the Einstein-Rosen bridge in the effective AdS$_2$ spacetime \eqref{eq:weylliou}:
\begin{equation}\label{LL}
L = \int_\mathcal{\text{geodesic}}\mathrm{d} s_{\text{eff}} = \int_\mathcal{\text{geodesic}} \mathrm{d} s_{\text{flat}} \,e^{b\varphi}.
\end{equation}
where $\varphi$ is the Liouville field in the original Liouville gravity formulation.
In this language, the boundary two-point function (which encoded the quantum disk isometries) is an insertion of the operator $e^{- \Delta L}$ 
in the $q$-Liouville path integral \cite{Blommaert:2023wad}. Therefore, it is this Liouville metric that encodes the non-commutative features of the quantum disk in the probe regime. We can then suggestively interpret the situation as follows. Since the physical dilaton gravity geometry is given by the geometry in the Liouville frame $\varphi$ up to a Weyl rescaling, this suggests also the dilaton gravity geometry is becoming fuzzy. Hence the classical singularity at the boundary \eqref{R} is fuzzy, and it then seems resolved quantum mechanically by the non-commutative geometry of the quantum hyperbolic disk.\footnote{In the past, non-commutative geometry has been indeed proposed as one of the main mechanisms to resolve curvature singularities.} From both perspectives (the dilaton gravity metric \eqref{geom} or the non-commutative hyperbolic disk), the conclusion would be that quantum fluctuations are unsuppressed near the holographic boundary.

\begin{figure}[h]
    \centering
    \includegraphics[width=0.55\linewidth]{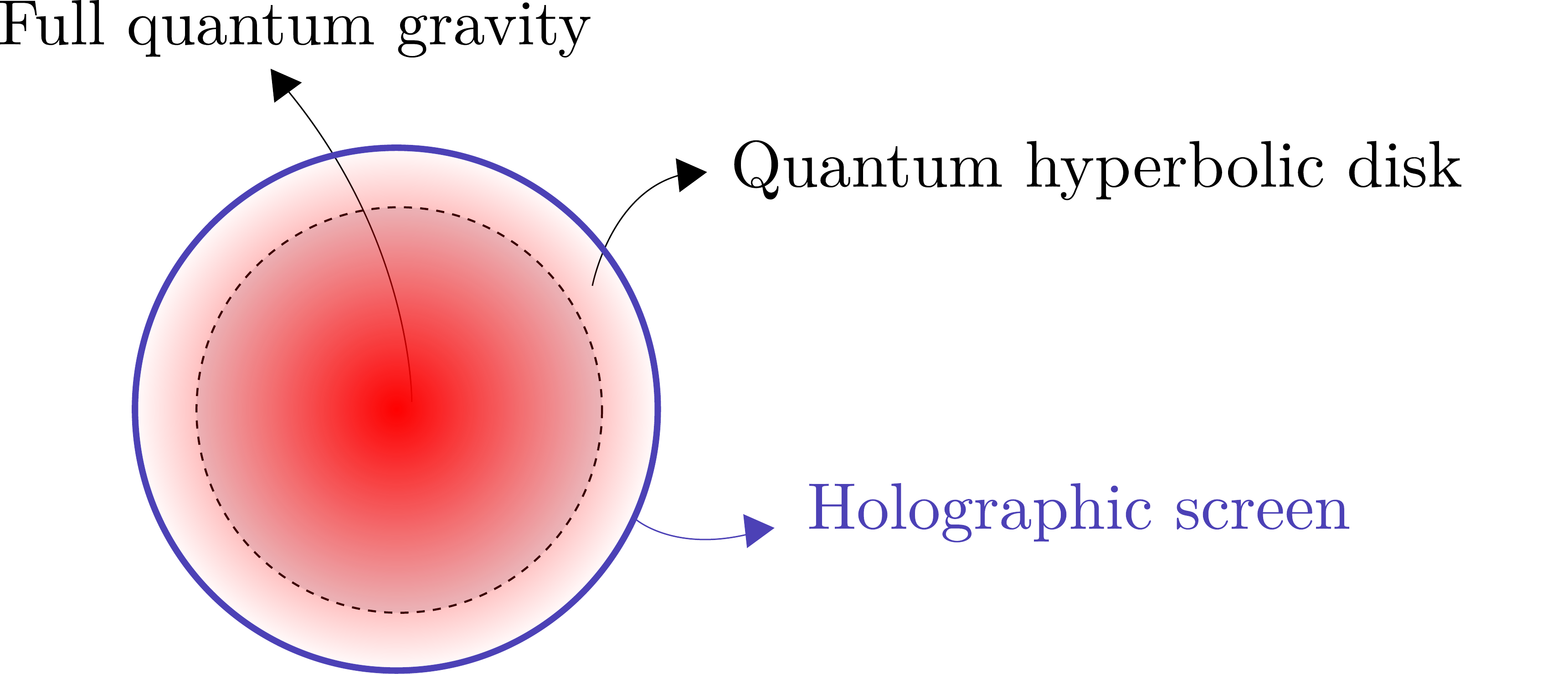}
    \caption{Geometry of the quantum sinh dilaton gravity bulk, with a full quantum gravity region deep in the interior, and a UV region closer to the holographic screen where we have an emergent quantum hyperbolic disk and SL$_q(2,\mathbb{R})$ symmetry. One way to phrase this is that close to the boundary the gravitational quantum fluctuations are reduced from full quantum gravity to just those describing the non-commutativity of the quantum hyperbolic disk spacetime.}
    \label{Figfinal}
\end{figure}

We summarize the final bulk structure of the \emph{quantum} model in Fig.~\ref{Figfinal}.
To get a more quantitative estimate on the extent of this region, we can superimpose the classical metric solution \eqref{geom}, and use those coordinates to delineate the region of interest. Since $r_h = \Phi_h$ and $E \sim \cosh 2\pi b^2 \Phi_h$, we can see that the large energy regime corresponds to where $\cosh 2 \pi b^2 r_h \approx e^{2 \pi b^2 r_h}/2$. The region outside of the classical black hole $r>r_h$ is then approximated by the metric
\begin{equation}
\mathrm{d} s^2 \approx - \frac{1}{4\pi^2b^4} \left[e^{2\pi b^2 r} - e^{2 \pi b^2 r_h} \right] dt^2 + \frac{4\pi^2b^4 dr^2}{ \left[e^{2\pi b^2 r} - e^{2\pi b^2 r_h} \right]},
\end{equation}
and identified with where we can replace cosh(.) with exp(.) in the energy integrals in \eqref{eq:start}, and hence where the quantum hyperbolic disk regime holds. We conclude that the quantum hyperbolic disk region is the asymptotic region of the bulk where the dimensionless radial coordinate $r$ of the classical solution satisfies
\begin{equation}
r \gg 1/b^2.
\end{equation}

\section{Concluding remarks}
\label{s:concl}
In this work, we have investigated the probe regime in a $q$-deformed setting of holography. We then illustrated that sinh (or sine) dilaton gravity exhibits this structure explicitly close to its holographic screen. There is an priori puzzling discrepancy between the semi-classical regime and the probe regime, which we interpret as due to there being a curvature singularity at the boundary. This singularity gets resolved by the non-commutative geometry of the quantum hyperbolic disk. We end by listing some possible directions for the future. \\

\textit{Other boundary correlation functions in the probe regime} \\
It should in principle be possible to apply our probe calculational method to other sinh dilaton gravity boundary correlation functions. In particular, the boundary three-point function and bulk-boundary two-point function have been explicitly computed in \cite{Mertens:2020hbs}. It would be interesting to explicitly take the probe limit in those cases and match it in the case of the boundary three-point function with known expressions for the $q$-holography boundary correlator. The bulk-boundary correlator would be an interesting new addition. \\

\textit{Supersymmetric generalization} \\
Another immediate possibility would be to consider the analogous dilaton supergravity models. The key player in our analysis was Ramanujan's formula \eqref{ram}. For the $\mathcal{N}=1$ sinh dilaton supergravity model for example, the relevant boundary two-point function correlators were determined in \cite{Fan:2021bwt}, and the relevant Ramanujan formula can be found in eq. (B.3) in \cite{Hadasz:2013bwa}. So it appear all of the ingredients are known to immediately pursue this generalization. \\

\begin{figure}[h]
    \centering
    \includegraphics[width=0.3\linewidth]{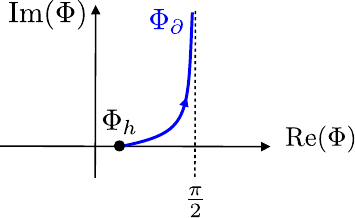}
    \caption{Complexified metric / dilaton contour for sine dilaton gravity, with the boundary at $\Phi_\partial = \frac{\pi}{2} + i\infty$.}
    \label{Figcomplcont}
\end{figure}

\textit{DSSYK and sine dilaton gravity} \\
It is important to try to adapt our techniques to the case of sine dilaton gravity, since the holographic boundary has been proposed to be of the same nature as the current sinh dilaton gravity case, but in a complexified direction of the geometry (Fig.~\ref{Figcomplcont}). 
Close to this complexified $\Phi_\partial$ along this contour, the geometry looks the same as the near-boundary sinh dilaton geometry we discussed in this work, and this is hence where we would expect the quantum hyperbolic disk to live. 
To implement our strategy concretely, first of all we need the technical Ramanujan identity.
Additionally we need to properly define the probe regime in that setting since the energy spectrum is bounded from above, so defining the UV-regime is more subtle. \\

\textit{Towards $q$-Virasoro as the asymptotic symmetry group} \\
Our holographic screen exhibits the SL$_q(2,\mathbb{R})$ isometry. However, just as the global $\mathfrak{sl}(2,\mathbb{R})$ algebra in CFT$_1$ is part of the much larger Virasoro algebra, also here the quantum algebra $U_q(\mathfrak{sl}(2,\mathbb{R})$ is to be embedded in a $q$-deformation of the Virasoro algebra, playing the role of the full asymptotic symmetry group. This structure has been defined and studied in the past in e.g. \cite{Shiraishi:1995rp,Awata:1995zk,Nieri:2013yra,Nieri:2013vba,Nedelin:2016gwu}. It would be interesting to understand this aspect of $q$-holography more explicitly within the context of our work. \\

\begin{figure}[h]
    \centering
    \includegraphics[width=0.25\linewidth]{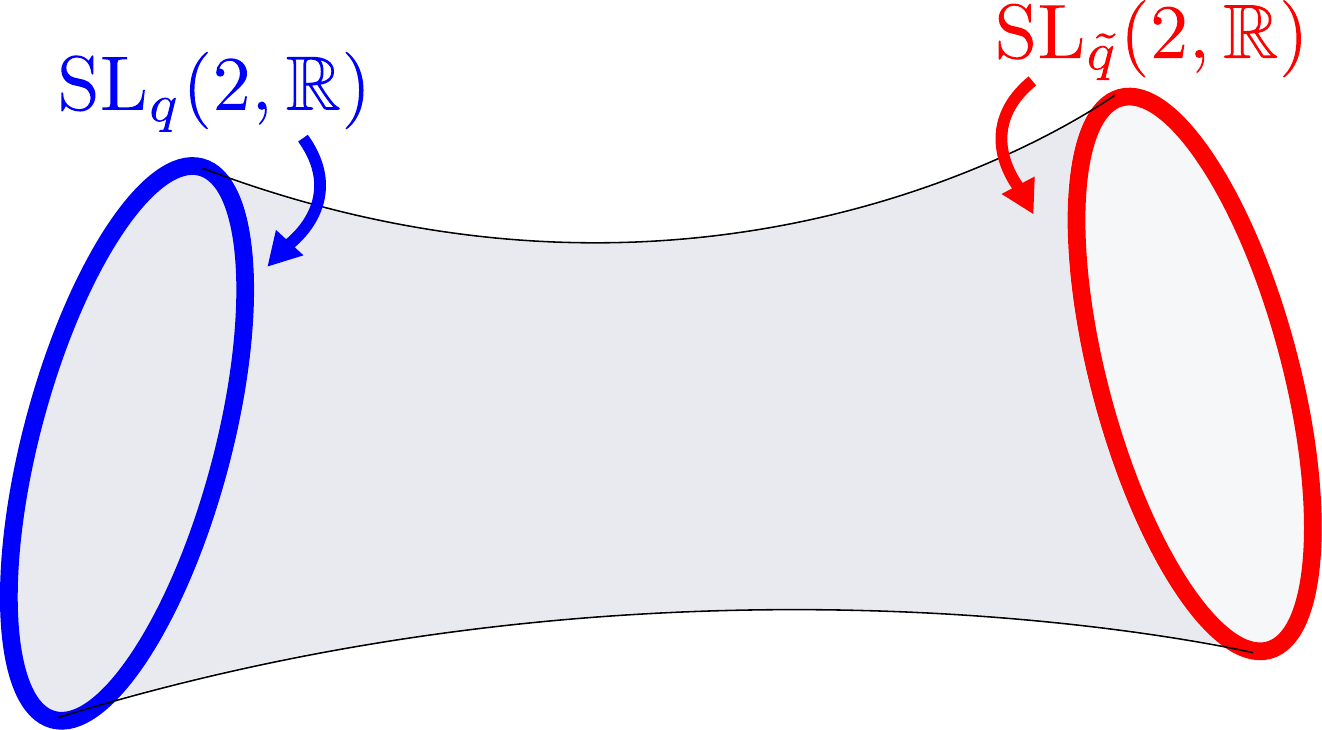}
    \caption{Example of a two-boundary amplitude, where the two boundaries ``geometrize'' distinct quantum hyperbolic disk isometries: SL$_q(2,\mathbb{R})$ versus SL$_{\tilde{q}}(2,\mathbb{R})$.}
    \label{Fig2dualbdy}
\end{figure}

\textit{Modular double and dual boundaries} \\
Our $q$-Ward identities isolated only half of the full symmetry algebra that governs sinh dilaton gravity. Indeed, its quantization is governed by the modular double $U_q(\mathfrak{sl}(2,\mathbb{R}))\otimes U_{\tilde{q}}(\mathfrak{sl}(2,\mathbb{R}))$, which is a Hopf algebra generated by six (or five) elementary generators \cite{Kharchev:2001rs,Bytsko:2002br,Bytsko:2006ut}, instead of the three that we exhibited in our construction in Section \ref{s:qward}. Here $q=e^{\pi i b^2}$ and $\tilde{q} = e^{\pi i/b^2}$. One way of seeing the other three generators is as follows. It is known in Liouville CFT that there are actually two dual types of FZZT branes \cite{Fateev:2000ik,Teschner:2000md}, related by the quantum duality $b\to 1/b$. Practically, this corresponds to changing $b\to 1/b$ directly in the energy expression in e.g. \eqref{partfuncs}. Close to such a boundary circle, one hence finds the dual quantum disk to appear as the isometry group. This means that the near-boundary isometry group for every single boundary is only half of the full modular quantum group symmetry governing the full model. An intriguing example would then be to consider a multiboundary amplitude with different (dual) types of boundaries, exhibiting and ``geometrizing'' the different SL$_q(2,\mathbb{R})$'s in the modular double (see Fig.~\ref{Fig2dualbdy} for an illustration). 
\newline

\textit{Fake disk versus the Drinfeld element in quantum groups?} \\
The boundary correlator can be written as
\begin{equation}
\label{eq:lgr}
\text{Tr} \left[\mathcal{O}\left(\tau_1-\frac{i}{2}\ln q\right)\mathcal{O}\left(\tau_2+\frac{i}{2}\ln q\right)e^{-\beta H}\right],
\end{equation}
where we wrote the sinh dilaton gravity boundary operators in terms of new operators $\mathcal{O}(\tau)$ whose correlators directly solve the $q$-Ward identities. Since time translation in $\tau$ is generated by the generator $K$ in the quantum algebra, we can rewrite \eqref{eq:lgr} as
\begin{align}
\text{Tr} \left[e^{-\frac{i}{2}\ln q H}\mathcal{O}(\tau_1)e^{i\ln q H}\mathcal{O}(\tau_2)e^{-\frac{i}{2}\ln q H}e^{-\beta H}\right]
&= \text{Tr} \left[\mathcal{O}(\tau_1)K^{-1/2}\mathcal{O}(\tau_2)e^{-\tilde{\beta} H}\right] \\
&= \text{Tr}_q \left[\mathcal{O}(\tau_2)e^{-\tilde{\beta} H}\mathcal{O}(\tau_1)\right],
\end{align}
where $\tilde{\beta} \equiv \beta - i\ln q = \beta +\pi b^2$, a slightly longer thermal circle.
The $K^{-1/2}$ element in the middle is interpretable as a defect insertion here. In the last line, we have defined a quantum trace by $\text{Tr}_q(.) \equiv \text{Tr}(K^{-1/2}.)$. The non-trivial observation here is that a Drinfeld-like element in the quantum trace appears naturally here from the fake disk extra length segment. We leave a more detailed investigation to future work.

\section*{Acknowledgments}
We thank A. Almheiri, P. Betzios, A. Blommaert, T. Tappeiner, B. Torres, G. Wong, Q.-F. Wu, and S. Yao for discussions. AB acknowledges the UGent Special Research Fund (BOF) for financial support. TM and JP acknowledges financial support from the European Research Council (grant BHHQG-101040024). Funded by the European Union. Views and opinions expressed are however those of the author(s) only and do not necessarily reflect those of the European Union or the European Research Council. Neither the European Union nor the granting authority can be held responsible for them.

\appendix
\section{Corepresentations of (quasi)-primary operators}\label{app:cft}
Holomorphic primary operators $\mathcal{O}(z)$ with weight $\Delta$ in a 2d CFT are defined to transform under the conformal transformation $z\rightarrow f(z)$ as
\begin{align}
    \mathcal{O}(z)\rightarrow \left(\frac{\partial f}{\partial z}\right)^\Delta \;\mathcal{O}(f(z)).
\end{align}
The corresponding generators under the infinitesimal transformation $f(z)=z+\epsilon z^{n+1}$ are given by $L_n=\Delta(n+1) z^n+z^{n+1}\partial_z$. Under a global PSL$(2,\mathbb{R})$ Möbius transformation $z\rightarrow f(z)=\frac{az+c}{bz+d}$ with $g=\begin{pmatrix}
    a&b\\c&d
\end{pmatrix}\in \text{SL}(2,\mathbb{R})$, (quasi)-primary operators consequently transform as 
\begin{align}
\label{eq:principalseries}
    \mathcal{O}(z)\rightarrow \left(bz+d\right)^{-2\Delta} \;\mathcal{O}\left(\frac{az+c}{bz+d}\right),
\end{align}
using the defining determinant property $ad-bc=1$. This is equivalent to the SL$(2,\mathbb{R})$ principal series representation analytically continued to the complex plane. Under global Möbius transformations with $n=-1,0,1$, the restricted set of generators coincide with the form of the principal series Borel-Weil generators satisfying the $\mathfrak{sl}(2,\mathbb{R})$ algebra $[L_0,L_{\pm1}]=\pm L_{\pm1}, \;[L_{+1},L_{-1}]=-2L_0$: \begin{align}\label{eq:BorelWeil}
    L_{-1}=\partial_z, \qquad L_{0}=z\partial_z+\Delta , \qquad L_{+1}=z^2\partial_z+2\Delta z,
\end{align} 
with corresponding fundamental generators 
\begin{align}\label{eq:fundamentalgenerators}
    E^-=\begin{pmatrix}
        0&0\\1&0
    \end{pmatrix}, \qquad H=\frac{1}{2}\begin{pmatrix}
        1&0\\0&-1
    \end{pmatrix}, \qquad E^+=\begin{pmatrix}
        0&-1\\0&0
    \end{pmatrix}.
\end{align} 
The Gauss-Euler parametrization implements a pairing between these fundamental $\mathfrak{sl}(2,\mathbb{R})$ generators and the SL$(2,\mathbb{R})$ coordinates used in \eqref{eq:principalseries} as 
\begin{align}
    g=e^{\beta E^-}e^{2\phi H }e^{\gamma E^+}=\begin{pmatrix}
        e^{\phi}&-\gamma e^\phi\\ \beta e^\phi&e^{-\phi}-\gamma\beta e^\phi
    \end{pmatrix}.
\end{align}
The infinitesimal action of \eqref{eq:principalseries} under these coordinates reproduces the respective principal series generators \eqref{eq:BorelWeil}. Correlation functions should be invariant under the global SL$(2,\mathbb{R})$ conformal subgroup, which is implemented by the three Ward identities 
\begin{align}
    L_n\braket{\mathcal{O}(z_1)\dots \mathcal{O}(z_N)}=m\circ\braket{\Delta^{(N)}(L_n)[\mathcal{O}(z_1)\otimes\dots \otimes\mathcal{O}(z_N)]}\equiv 0, \quad n=-1,0,+1,
\end{align}
where we have defined a linear multiplication map $m:\text{SL}(2,\mathbb{R})\otimes \text{SL}(2,\mathbb{R})\rightarrow \text{SL}(2,\mathbb{R})$ which multiplies both entries of the tensor product. The coproduct $\Delta:\text{SL}(2,\mathbb{R})\rightarrow \text{SL}(2,\mathbb{R})\otimes \text{SL}(2,\mathbb{R})$ trivially distributes the action of the generators on the different fields  
\begin{align}
    \Delta(L_n)=L_n\otimes \mathbf{1}+\mathbf{1}\otimes L_n.
\end{align}
By induction one can easily extend the coproduct to $\text{SL}(2,\mathbb{R})\rightarrow (\text{SL}(2,\mathbb{R}))^{\otimes N}$ as 
\begin{align}
\label{eq:copN}
    \Delta^{(N)}(L_n)=L_n\otimes \mathbf{1}\otimes \dots \otimes \mathbf{1}\;+\;\mathbf{1}\otimes L_n\otimes \dots \otimes \mathbf{1}\;+\;\dots\;+\;\mathbf{1}\otimes \dots \otimes \mathbf{1}\otimes L_n,
\end{align}
yielding a trivial Leibniz rule, making the operators in the correlation function cyclically symmetric. 

To define the global SL$_q(2)$ subgroup of a $q$-deformed conformal field theory, we first define the larger quantum group GL$_q(2)$ \cite{Ip:2013}. This quantum group can naturally be defined as the algebra generated by the non-commutative variables $z_{ij},\,\, i,j = 1,2$, subject to the commutation relations
\begin{gather}
\left[ z_{11},  z_{12} \right] = \left[ z_{21},  z_{22} \right] = 0, \qquad  
\left[ z_{12},  z_{21} \right]  = \left[ z_{11},  z_{22} \right], \\
z_{11}z_{21} = q^2 z_{21} z_{11},  \qquad z_{12}z_{22} = q^2 z_{22}z_{12}, \qquad 
z_{12}z_{21} = q^2 z_{21} z_{12},
\end{gather}
where we can view $z_{ij}$ as elements of a fundamental $2 \times 2$ matrix and the coproduct is realized as a matrix product 
\begin{align}
    \Delta(z_{11})&=z_{11}\otimes z_{11}+z_{12}\otimes z_{21}, \qquad \Delta(z_{12})=z_{11}\otimes z_{12}+z_{12}\otimes z_{22},\\ \Delta(z_{21})&=z_{21}\otimes z_{11}+z_{22}\otimes z_{21},\qquad \Delta(z_{22})=z_{21}\otimes z_{12}+z_{22}\otimes z_{22}. 
\end{align} 
We further note that the determinant of this matrix $N \equiv z_{11}z_{22} - z_{12} z_{21}$ satisfies 
\begin{equation}
    Nz_{ii} = z_{ii}N, \qquad N  z_{12} = q^{-2} z_{12} N, \qquad N z_{21} = q^{2} z_{21}  N.
    \label{eq:commN}
\end{equation} 
Using the matrix coproduct relations, one easily gets $\Delta(N)=N\otimes N$.

This defines the GL$_q(2)$ quantum group as a Hopf algebra of coordinates. The restriction to $\text{SL}_q(2)$ is then implemented via the identification \cite{Ip:2013}
\begin{equation}
\label{eq:projection}
    \begin{pmatrix}
        a & b \\ c & d
    \end{pmatrix} \equiv  N^{-1/2}\begin{pmatrix}
z_{11} & q^{-1/2}z_{12} \\q^{1/2} z_{21} & z_{22}        
    \end{pmatrix},
\end{equation}
whose coordinates satisfy the defining SL$_q(2)$ commutation relations 
\begin{align}
\label{eq:coordinatealgebra}
   ab = qba, \quad cd = qdc, \quad ac = qca, \quad bd &= qdb, \quad ad - da = (q - q^{-1})bc \\
    \det{}_q &\equiv ad - qbc = 1.
\end{align}
The last relation defines a notion of a $q$-determinant $\det\{\}_q$, which is central, and restricted to one in the definition of the coordinate Hopf algebra. In particular, we arrive at the defining SL$_q(2)$ determinant constraint by multiplying the GL$_q(2)$ matrices by the inverse square root of the determinant operator.

By analogy with the transformation rule of (quasi)-primary operators in a classical CFT \eqref{eq:quasiprimarytransf}, we define (quasi)-primary operators in $q$-deformed conformal field theory to transform under the global SL$_q(2)$ subgroup in terms of a ``finite'' principal series corepresentation $\phi$, constructed analogously in \cite{Belaey:2025ijg}:  
\begin{align}
\label{eq:qprimary}
 \mathcal{O}(z) \mapsto \phi(\mathcal{O}(z)) \equiv  (1\otimes N^{\Delta/2}) \left( z\otimes z_{12} + \mathbf{1}\otimes z_{22} \right)^{-2\Delta} \;\mathcal{O}\left( \frac{z\otimes z_{11} + \mathbf{1}\otimes z_{21}}{ z\otimes z_{12} +\mathbf{1}\otimes z_{22}}\right) (\mathbf{1}\otimes N^{\Delta/2}),
\end{align} 
where we assume $\mathcal{O}(z)$ to be a sum of monomials. Noting that since $[z\otimes z_{11} + 1\otimes z_{21}, z\otimes z_{12} + 1\otimes z_{22}] = 0$, the expression is independent of ordering up to the determinant factors \cite{Ip:2013}. These determinant factors project down to a corepresentation of SL$_q(2)$ by multiplying the GL$_q(2)$ coordinates with the appropriate inverse square roots of the determinant via \eqref{eq:projection}.  A coprepresentation of the Hopf algebra SL$_q(2)$ is defined to act on a vector space $V$ with a linear map $\phi:V \to V\otimes \text{SL}_q(2)$ such that\footnote{The classical analogue of this property is the familiar factorization property $\phi(g_1\cdot g_2)=\phi(g_1)\phi(g_2).$}$^{,}$\footnote{Note that a corepresentation has the opposite arrow in the commutative diagram compared to a normal representation.} 
\begin{align}
    (\text{id} \otimes \Delta) \circ \phi =(\phi \otimes \text{id}) \circ \phi.
\end{align}
In particular, one can easily check that the transformation rule satisfies \begin{align}
     (&\phi \otimes \text{id})\left( (\mathbf{1}\otimes N^{\Delta/2}) \left( z\otimes z_{12} + \mathbf{1}\otimes z_{22} \right)^{-2\Delta} \mathcal{O}\left( \frac{z\otimes z_{11} +\mathbf{1}\otimes z_{21}}{ z\otimes z_{12} + \mathbf{1}\otimes z_{22}}\right) (\mathbf{1}\otimes N^{\Delta/2})\right)  \\
     &=  (\mathbf{1}\otimes \Delta(N)^{\Delta/2}) (z\otimes \Delta(z_{12}) + \mathbf{1}\otimes \Delta(z_{22}))^{-2\Delta}\mathcal{O}\left( \frac{z\otimes \Delta(z_{11}) + \mathbf{1}\otimes\Delta(z_{21})}{z\otimes \Delta(z_{12}) + \mathbf{1}\otimes\Delta(z_{22})} \right) (\mathbf{1}\otimes \Delta(N)^{\Delta/2}).\nonumber
\end{align}

Analogous to the undeformed CFT theory, infinitesimal transformations are implemented by an algebra of generators. In particular, we define the $U_q(\mathfrak{gl}_2)$ enveloping algebra of the generators $E^+,E^-,K,K_0$ subject to the algebra relations \begin{align}\label{eq:generatoralgebra}
    KE^\pm=q^{\pm 2}E^\pm K, \qquad KK^{-1}=K^{-1}K=1, \qquad [E^+,E^-]=-\frac{K-K^{-1}}{q-q^{-1}},
\end{align} and $K_0$ commuting with the other generators. This is a Hopf algebra under the coproduct 
\begin{align}
    \Delta(K) = K \otimes K, &\quad \Delta(E^+) = K_0 \otimes E^+ + E^+ \otimes K_0^{-1}K^{-1}, \\
    \Delta(K_0)= K_0 \otimes K_0,& \quad \Delta(E^-) =  K_0^{-1}K \otimes E^- + E^- \otimes K_0.
\end{align}  
One can project down to the $U_q(\mathfrak{sl}_2)$ Hopf algebra by setting the central generator $K_0$ to unity.
 Analogous to the classical generators in the fundamental representation \eqref{eq:fundamentalgenerators}, we choose a bilinear pairing between the fundamental generators $E^-,E^+,K,K_0$ of the enveloping algebra $U_q(\mathfrak{gl}_2)$ and the coordinates of GL$_q(2)$, $\braket{\cdot,\cdot}: U_q(\mathfrak{gl}(2)) \times \text{GL}_q(2) \to \mathbb{C}$ by the non-vanishing brackets:
\begin{align}
\label{eq:pairings}
    \braket{K,z_{11}} = q, &\quad \braket{K,z_{22}} = q^{-1},\\
    \braket{K_0,z_{11}} = q^{-1/2}, &\quad \braket{K_0,z_{22}} = q^{-1/2},\\
    \braket{E^+,z_{12}} = -1, &\quad \braket{E^-,z_{21}} = 1.  
\end{align} 
The pairing is extended to multiplicative words of the enveloping algebra by demanding that for every $X,Y\in U_q(\mathfrak{gl}_2)$ and $g_1,g_2\in \text{GL}_q(2)$ 
\begin{align}
    \braket{X,g_1 g_2} &= \braket{\Delta(X),g_1 \otimes g_2}, \qquad
    \braket{XY,g} = \braket{X \otimes Y, \Delta(g)}.
    \label{eq:hopfduality}
\end{align}
In this way, the Hopf algebra of the generators $U_q(\mathfrak{gl}_2)$ is dual to the Hopf algebra of coordinates GL$_q(2)$. 

Using this duality between the two Hopf algebras, the transformation rule of the (quasi)-primary operators under SL$_q(2)$ \eqref{eq:qprimary} implies a representation of the infinitesimal generators acting as operators on functions on the complex plane, yielding the analogue of the classical ``infinitesimal'' generators \eqref{eq:BorelWeil}. In particular, given the form of the corepresentation \eqref{eq:qprimary} $\phi : V \to V \otimes \text{SL}_q(2)$ and a pairing of Hopf algebras $\braket{\cdot, \cdot} : U_q(\mathfrak{sl}_2) \times \text{SL}(2) \to \mathbb{C}$ we can induce a representation $\hat{\phi}$ of $U_q(\mathfrak{sl}_2)$ on $V$ by
\begin{equation}
    \hat{\phi}(X) v \equiv (\text{id} \otimes \braket{X, \cdot}) \circ (\phi(g) v),\qquad v\in V, \quad g\in \text{SL}_q(2), \quad X\in U_q(\mathfrak{sl}_2).
\end{equation}
Introducing a scaling operator $R_{q^2}f(z)\equiv f(q^2z)$, we calculate the principal series generators from the above pairings between the generators and coordinates as:\footnote{The calculation was done explicitly in appendix A of \cite{Belaey:2025ijg} for the symmetric version of the principal series representation, but can easily be extended to the case at hand.} 
\begin{align}
\label{eq:qprincipalgenerators}
    K=q^{2\Delta}R_{q^2}, \qquad E^-=\frac{q^{-1/2+\Delta}}{z}\frac{R_{q^2}-1}{q-q^{-1}}, \qquad E^+=-q^{-1/2+\Delta} z\frac{q^{-4\Delta} R_{q^{-2}}-1}{q-q^{-1}}.
\end{align}
These are the familiar principal series generators of $U_q(\mathfrak{sl}_2)$ \cite{Ponsot:1999uf,Ponsot:2000mt,Kharchev:2001rs} used e.g. to describe the amplitudes of Liouville gravity. One can check explicitly that these generators satisfy the correct U$_q(\mathfrak{sl}_2)$ algebra relations \eqref{eq:generatoralgebra}. The inclusion of the determinant factors in the SL$_q(2)$ principal series corepresentation \eqref{eq:qprimary} ensures that the dual $K_0$ generator acts trivially as the identity operator, effectively projecting down to the subalgebra $U_q(\mathfrak{sl}_2)$. Introducing a $q$-derivative operator $D_{q}=\frac{R_q-1}{x(q-1)}$ and a $q$-number $[x]_q=\frac{q^{x}-1}{q-1}$, we can suggestively rewrite these operators as 
\begin{align}
    K=q^{2\Delta}R_{q^2}, \quad E^-=q^{1/2+\Delta}D_{q^2}, \qquad E^+=q^{-3/2-3\Delta}z^2D_{q^{-2}}+q^{-3/2+\Delta}z[2\Delta]_{q^{-2}},
\end{align}  
which in the classical $q\rightarrow1$ limit manifestly yield the infinitesimal principal series generators of the non-deformed CFT \eqref{eq:BorelWeil} with the identification $K\equiv q^{2H}$.

\section{Dilaton gravity models}
\label{app:dil}
The bulk action of a generic (Euclidean) dilaton gravity model with potential $V(\Phi)$ is given by
\begin{align}
    S=-\frac{1}{2}\int d^2x \sqrt{g}\left(\Phi R+V(\Phi)\right).
\end{align} 
One may gauge-fix the classical solutions to be of the form\footnote{In the asymptotic region, the general classical solution is actually $\Phi(r) \approx a r$. The prefactor $a$ is here set to 1. Note that there is physics contained in $a$, since the slope with which the dilaton diverges at the boundary is part of the definition of the boundary conditions, just like in JT gravity where this prefactor is essentially the Schwarzian coupling coefficient, usually denoted by $C$ \cite{Mertens:2022irh}.} 
\begin{align}
    ds^2=F(r) d\tau^2 +\frac{dr^2}{F(r)}, \qquad \Phi=r.
\end{align}
In this gauge, we have $R=-F''$, such that the solution to the equations of motion is given by 
\begin{align}
    F(r)=\int_{r_h}^r dr'\; V(r'),
\end{align}
where $r_h$ is the radius of the horizon for which the metric vanishes at fixed radius. Expanding the metric around the horizon and demanding regular periodicity in Euclidean time, leads a relation between the temperature and potential 
\begin{align}\label{eq:temp}
    \beta = \frac{4\pi }{\abs{V(r_h)}}.
\end{align}
From the BH entropy formula 
\begin{equation}
S_{BH} = 2\pi \Phi_h,
\end{equation}
we then obtain from the first law $dE=TdS$:
\begin{equation}
E = \frac{1}{2}\int^{V^{-1}(4\pi T)}V(\Phi)d\Phi.
\end{equation}
This will be reproduced below from an on-shell action approach.

\subsection{Holographic counterterm}
\label{a:counter}

The GHY boundary action with appropriate counter term is of the form
\begin{equation}
- \oint_{\partial \mathcal{M}} d \tau\,\sqrt{h}\,\bigg(\Phi K-\left(\int^{\Phi}_{c} V(\Phi)d\Phi\right)^{1/2} \bigg),
\end{equation}
where the lower bound $c$ of the counterterm integral is arbitrary and just corresponds to an energy offset.\footnote{This specific boundary term was used in the case of sinh dilaton gravity in \cite{Blommaert:2023wad} and for sine dilaton gravity in \cite{Blommaert:2024ymv}. Here we compile the more general story. The argument follows largely the computation presented in \cite{Witten:2020ert} for the case of (deformations of) JT gravity.} We can motivate this counterterm explicitly by looking at the full on-shell Euclidean action
\begin{equation}
S_E = -\frac{1}{2} \int_\mathcal{M} d^2x \sqrt{g}\left(\Phi R + V(\Phi)\right) - \oint_{\partial \mathcal{M}} d \tau\,\sqrt{h}\,\bigg(\Phi K-\sqrt{\int^{\Phi_{\partial}}_c V(\Phi) d\Phi }\bigg).
\end{equation}
Using $R = -V'(\Phi)$ and denoting $F(r) = \int_{r_h}^{r}dr \, V(r)$, the bulk on-shell action on the classical solution in ``Schwarzschild gauge'' leads to the following expression
\begin{align}
S_{b}^{\text{on-shell}} &= -\frac{\beta}{2}\int_{r_h}^{r_{\partial}} dr (-r F''+F') = -\frac{\beta}{2} [-rF'+2F]_{r_h}^{r_\partial} \\
&= -\frac{\beta}{2}[{\color{darkblue}-r_\partial V(r_\partial)} + 2F(r_\partial) +r_h V(r_h)].
\end{align}
The GHY boundary term + counterterm on the other hand gives:
\begin{align}
S_{\partial}^{\text{on-shell}} &= -\frac{\beta}{2} \left[{\color{darkblue}r_{\partial}V(r_\partial)} - 2\sqrt{F(r_\partial)}\left(\int^{\Phi}_{c} V(\Phi)d\Phi\right)^{1/2}\right].
\end{align}
The GHY term (in darkblue) cancels a corresponding divergent bulk contribution, and the total on-shell action reduces to
\begin{align}
S_{b}^{\text{on-shell}} + S_{\partial}^{\text{on-shell}} = -2\pi \Phi_h -\beta \left(F(r_\partial) -\sqrt{F(r_\partial)}\left(\int^{\Phi}_{c} V(\Phi)d\Phi\right)^{1/2}\right),
\end{align}
where we used $V(r_h) = 4\pi/\beta$ to disentangle the entropy contribution and the energy term. 

The boundary counterterm is the unique option that only depends on the fields (and not on properties of a specific solution).
The resulting final expression for the ADM energy is then suggestively written as
\begin{equation}
E_{\text{ADM}} = - \int^{\Phi_{\partial}}_{\Phi_h} V(\Phi) d\Phi + \sqrt{\int^{\Phi_{\partial}}_{\Phi_h} V(\Phi) d\Phi \times \int^{\Phi_{\partial}}_c V(\Phi) d\Phi} \,\, \approx \,\, \frac{1}{2}\int^{\Phi_h}_{c}V(\Phi)d\Phi,
\end{equation}
under the assumption that $\Phi_{\partial}$ is such that $\int^{\Phi_{\partial}} V(\Phi) d\Phi \to +\infty$.
The first term here comes from the bulk on-shell term, and the second term coming from the boundary counterterm. This result is precisely the known expression for the ADM energy of a general dilaton gravity model. 

For the specific case of the sinh (and sin) dilaton potentials, one can check that the boundary counterterm can be equivalently written explicitly in various guises:
\begin{equation}
\label{eq:countervari}
\oint_{\partial \mathcal{M}} d \tau\,\sqrt{h}\,\sqrt{\frac{\cosh(2\pi b^2 \Phi)-1}{2\pi^2 b^4}} =  \oint_{\partial \mathcal{M}} d \tau\,\sqrt{h}\,\sqrt{\frac{\cosh(2\pi b^2 \Phi)}{2\pi^2 b^4}} = \oint_{\partial \mathcal{M}} d \tau\,\sqrt{h}\,\frac{\exp(\pi b^2 \Phi)}{2\pi b^2},
\end{equation}
which are equivalent since we have to take the leading term in the asymptotic expansion for $\Phi_\partial \to +\infty$. 

For the leftmost expression of \eqref{eq:countervari} (which we used in the main text \eqref{Liouville}), the $-1$ corresponds to the choice of integration constant $c=0$ above. This is done to precisely reproduce JT gravity in the $b\to 0$ limit.\footnote{There is an order of limits issue here, since to find JT we have to take $b\to 0$ before taking $\Phi_\partial \to +\infty$, corresponding to the near-boundary UV region of sinh dilaton gravity deviating from the JT AdS$_2$ spacetime as discussed in subsection \ref{s:classical}.}

The rightmost form \eqref{eq:countervari} on the other hand is natural when obtaining this system from the perspective of two Liouville actions, and can be readily shown to lead to the same on-shell energy expression:
\begin{equation}\label{eq:ADM}
E_{\text{ADM}} = \frac{1}{2}\int^{\Phi_h}_{c}V(\Phi)d\Phi,
\end{equation}
and hence the same physics as the other two ways of writing the boundary term.

As an aside, in the first-order formulation of dilaton gravity, the holographic boundary term is directly chosen to be the Casimir of the Poisson algebra (spanned by 3 generators $(\Phi_0,\Phi_1,\Phi_\partial)$) as 
\begin{equation}
\frac{1}{2}\oint_{\partial \mathcal{M}} d\tau \bigg(\Phi_0^2+\Phi_1^2 + \int^{\Phi_{\partial}}_c V(\Phi) d\Phi\bigg).
\end{equation}
The on-shell energy computation would now instead be
\begin{equation}
E_{\text{ADM}} = - \int^{\Phi_{\partial}}_{\Phi_h} V(\Phi) d\Phi + \frac{1}{2}\int^{\Phi_{\partial}}_{\Phi_h} V(\Phi) d\Phi + \frac{1}{2} \int^{\Phi_{\partial}}_c V(\Phi) d\Phi  = \frac{1}{2}\int^{\Phi_h}_{c}V(\Phi)d\Phi,
\end{equation}
yielding indeed the same expression for the energy.

\subsection{Sinh dilaton gravity quantization}
\label{a:quant}
In order to be self-contained, we follow here a more direct approach based on the canonical phase space quantization of the theory, which will allow us to directly identify the boundary dynamics and the associated symmetry algebra. A similar analysis was carried out for sine dilaton gravity in \cite{Blommaert:2024ymv,Blommaert:2024whf}, so we will not repeat the full set of equations here, but rather summarize the key steps leading to the derivation of the gravitational Hamiltonian for $\sinh$ dilaton gravity. 

The above formulation of the action \eqref{Liouville} is convenient to directly take the JT limit as $b\rightarrow0$. Here it is convenient to rescale $\pi b^2\Phi\rightarrow \Phi$ and take the latter to be $\Phi\sim 1$, such that the whole action is multiplied by $1/b^2$, and the $b$ parameter governs the semiclassical saddle as $\hbar\sim  b^2$: 
\begin{equation}
\label{scaledLiouville}
S = -\frac{1}{\pi b^2}\bigg( \frac{1}{2}\int_\mathcal{M} d^2x \sqrt{g}\left(\Phi R + \sinh (2 \Phi)\right) + \oint_{\partial \mathcal{M}} d \tau\,\sqrt{h}\,\bigg(\Phi K-\sqrt{\frac{\cosh(2 \Phi)-1}{2}}\bigg)\bigg).
\end{equation} 
This leads to an effective rescaled dilaton potential $V(\Phi)\rightarrow \pi b^2 V(\Phi)$. The corresponding rescaled energy and temperature can be deduced from the generic thermodynamic relations
\begin{align}
    E= \frac{\cosh(2 \Phi_h)}{4 \pi b^2}, \qquad \beta=\frac{4\pi}{\sinh(2\Phi_h)}.
\end{align}

In performing canonical quantization of general dilaton gravities, an apparent difficulty arises from the fact that there is no unique geodesic connecting two arbitrary boundary points \cite{Kruthoff:2024gxc}. To address this, in our case it is more natural to consider geodesic distances measured in a Weyl-rescaled geometry $\mathrm{d}s_{\mathrm{eff}}^2=\mathrm{d}s^2 e^{-2 \Phi}$. Writing the metric in conformal gauge as $\mathrm{d}s^2 = \mathrm{d}s_{\mathrm{flat}}^2 e^{2\rho}$, this effective metric 
\begin{equation}
\mathrm{d}s_{\mathrm{eff}}^2=\mathrm{d}s^2 e^{-2 \Phi}=\mathrm{d}s_{\mathrm{flat}}^2 e^{2\rho-2 \Phi}=\mathrm{d}s_{\mathrm{flat}}^2 e^{2b \varphi},
\end{equation} 
corresponds to the one associated with the Liouville field $\varphi$ in the original Liouville gravity formulation. Indeed the Liouville equation admits AdS$_2$ conformal factors as classical solutions. This leads to an effective AdS$_2$ black hole geometry in which the renormalized geodesic length $L$ between the two boundaries is well-defined:
\begin{equation}\label{L}
L = \int_\mathcal{\text{geodesic}}\mathrm{d} s_{\text{eff}} = \int_\mathcal{\text{geodesic}} \mathrm{d} s_{\text{flat}} \,e^{b\varphi}.
\end{equation}
A canonical transformation is then performed in phase space, replacing the original variables, namely the ADM energy $E_{\mathrm{ADM}}$ and its conjugate two-sided boundary time coordinate $T$, with a new pair: the renormalized Einstein-Rosen bridge length $L$ \eqref{L} and its conjugate momentum $P$. Specifically, one requires the canonical symplectic form in the original coordinates to be of the form $\omega= \mathrm{d}T\land \mathrm{d} E_{\mathrm{ADM}}$. Upon rescaling the coordinates to obtain a nice saddle, we have
\begin{equation}\label{omega}
\omega=\mathrm{d}T \wedge \mathrm{d}E_{\mathrm{ADM}}=\frac{\sinh(2\Phi_h)}{2\pi b^2}\mathrm{d}T\wedge \mathrm{d}\Phi_h=\frac{1}{  2\pi b^2} \mathrm{d}P \wedge \mathrm{d}L,
\end{equation}
such that after quantization, $\pi b^2$ takes the role of $\hbar$.
One can easily prove that the canonical transformations which ensure \eqref{omega} are
\begin{equation}\label{rel}
\begin{split}
L&=2\log \left(\frac{\cosh \left(T \sinh (2 \Phi_h)/4\right)}{\sinh (2 \Phi_h)}\right),\\ 
P&=-\log \left[\cosh(2 \Phi_h)+\sinh (2 \Phi_h)\tanh \left(T \sinh (2 \Phi_h)/4\right)\right],
\end{split}
\end{equation}
where $L$ is indeed the length of a Cauchy slice in a two-sided AdS$_2$ black hole at the same inverse temperature as the original sinh dilaton gravity black hole.

We now invert the relations \eqref{rel} and express the gravity Hamiltonian \eqref{adm} in terms of the new phase space variables $L$ and $P$, obtaining:
\begin{equation}
\label{adm2}
4\pi b^2 H_{\mathrm{ADM}}= \cosh (P)+\frac{e^{-L}e^{P}}{2}.
\end{equation}
A remarkable feature is that, by performing a Legendre transform of the Hamiltonian \eqref{adm2}, one obtains the action of the so-called $q$-Liouville system. The latter is part of a more general theory, consisting of three coordinates $x_{a}=\left(L/2,\beta,\gamma\right)$ and three corresponding momenta $p_{a}=\left(P,p_{\beta},p_{\beta}\right)$, known as the $q$-Schwarzian \cite{Blommaert:2023wad,Blommaert:2023opb}. This six-dimensional phase space reduces to the two-dimensional phase space of the system \eqref{adm2} upon imposing two constraints, which were interpreted as Brown–Henneaux boundary conditions at the two asymptotic boundaries of the two-sided gravitational system \cite{Blommaert:2023wad,Blommaert:2023opb}. The $q$-Schwarzian system has classically 6 conserved charges which, upon quantization, satisfy two copies of the SL$_q(2)$ algebra \eqref{eq:generatoralgebra}. 
In conclusion, by showing that the boundary dynamics of Liouville gravity reduce to the constrained version \eqref{adm2} of the $q$-Schwarzian, which is basically the simplest dynamical system encoding the desired symmetry algebra, we argue that Liouville gravity may serve as a natural bulk gravitational dual capable of reproducing the two-point function obtained in the main text.

All amplitudes in sinh dilaton gravity, and in particular the matter two-point function we are interested in, can be obtained by quantizing the system \eqref{adm2}. From the symplectic form \eqref{omega}, the associated Poisson brackets $\{L,P\}=2\pi b^2$, become upon quantization $[L,P]=2\pi ib^2\hbar$. Setting $\hbar\equiv 1$ by absorbing it in the definition $\pi b^2\hbar\rightarrow \pi b^2$, we see that $b^2$ effectively governs the semiclassical regime. By realizing the canonical commutation relation by setting $P = -2 \pi i b^2 \frac{\mathrm{d}}{\mathrm{d}L}$, the Schrödinger equation associated with the Hamiltonian \eqref{adm2} becomes
\begin{equation}
\label{difference}
\psi_{s}(L/2+i \pi b^2)+\left(1+e^{i \pi b^2}e^{-L}\right)\psi_{s}(L/2-i \pi b^2)=2\cosh(2\pi b s)\psi_{s}(L/2),
\end{equation}
where, for convenience, we have introduced a new energy variables $\Phi_h=\pi bs$. The additional factor $e^{2\pi i b^2}$ is an order ambiguity in the quantization procedure. This difference equation appeared before within the context of the relativistic Toda chain \cite{Kharchev:2001rs}. The solution to this difference equation \eqref{difference} is known and becomes unique once we require it to also satisfy the dual equation obtained by the transformation $b \rightarrow 1/b$.\footnote{For completeness, the solution is given by \cite{Kharchev:2001rs,Mertens:2020hbs,Fan:2021bwt}
\begin{equation}
\psi_{s}(L/2) = e^{- i s L / b} \int_{\mathcal{C}} \mathrm{d}\xi \, S_{b}(-i \xi) S_{b}(-i \xi - 2i s) e^{-\pi i (\xi^2 + 2s \xi)} e^{- i \xi L / b},
\end{equation}
where $\mathcal{C}$ denotes an appropriate integration contour.} This requirement is not manifest at the level of the classical action, which is not invariant under this duality, but it emerges in the structure of the quantum amplitudes of Liouville gravity \cite{Mertens:2020hbs,Fan:2021bwt}.\footnote{Liouville gravity has a quantum group theoretic description in terms of the modular double $U_q(\mathfrak{sl}(2,\mathbb{R}))\otimes U_{\tilde{q}}(\mathfrak{sl}(2,\mathbb{R}))$ Hopf algebra \cite{Fan:2021bwt,Mertens:2022aou}, building on the same structure present in 2d Liouville CFT \cite{Ponsot:1999uf,Ponsot:2000mt}.} The boundary-to-boundary propagator of a matter scalar field in the bulk of sinh dilaton gravity is dual to the insertion of the operator 
$e^{- \Delta L}$ 
in the $q$-Liouville path integral \cite{Blommaert:2023wad}. Once the gravitational eigenfunctions are known, the matrix element of this operator between two fixed energy states 
$s_1$ and $s_2$ 
can be computed, yielding:
\begin{equation}\label{matrix}
\int_{-\infty}^{+\infty} \mathrm{d}\phi \ \psi_{s_1}(\phi)\, \psi_{s_2}^*(\phi)\, e^{-2 \Delta \phi} = \frac{S_b(b \Delta \pm i s_1 \pm i s_2)}{S_b(2b \Delta)},
\end{equation}
where the shorthand notation $a \pm b \pm c$ indicates the product over all combinations of signs. Taking the limit $\Delta \to 0$ the expression above reduces to the orthogonality relation between the eigenfunctions, from which one can extract the Liouville gravity density of states:
\begin{equation}
\label{spectral}
\rho(s) = \sinh(2\pi b s)\, \sinh\left(\frac{2\pi s}{b}\right).
\end{equation}
These two ingredients form the fundamental building blocks of the Liouville gravity two-point function~\eqref{two-point} and the partition function \eqref{partfuncs}.

\section{Useful manipulations involving Ramanujan's formula}
\label{app:form}
A crucial identity for this work is Ramanujan's summation formula:
\begin{align}\label{master}
\int_\mathbb{R} \mathrm{d} \tau e^{-2\pi \beta \tau} e^{\pi (Q-\alpha)\tau} e^{-\frac{i \pi}{2}\alpha(Q-\alpha)}e^{i\pi \alpha\beta} S_b(i\tau+\alpha)S_b(-i\tau) = \frac{S_b(\alpha)S_b(\beta)}{S_b(\alpha+\beta)}.
\end{align}
We now aim to rewrite it in a way that will be useful to us. 
The argument has poles of $S_b(-i\tau)$ at $\tau=-inb-im/b$ and at $\tau=i\alpha+inb+im/b$ for $S_b(i\tau+\alpha)$. The contour avoids the pole at $\tau=0$ from above. Deforming the contour to $\tau \to \tau - i \alpha/2$, we can rewrite this formula suggestively as
\begin{align}\label{Ramanujan}
\int_\mathbb{R} \mathrm{d}\tau e^{-2\pi \beta \tau} e^{\pi (Q-\alpha)\tau} \frac{S_b(\alpha/2 \pm i \tau)}{S_b(\alpha)}= \frac{S_b(\beta)}{S_b(\alpha+\beta)}.
\end{align}
Moreover, a useful rewriting of the exact Liouville gravity two-point function \eqref{two-point} can be achieved by exploiting the following version of Ramanujan's formula \eqref{Ramanujan}:
\begin{equation}\label{A.1}
\int_{-\infty}^{+\infty} \mathrm{d} p e^{2\pi ip x}  \frac{S_b(ip+Q/2-\beta_M)}{S_b(ip+Q/2+\beta_M)}= \frac{S_b(\beta_M \pm i x)}{S_b(2 \beta_M)}.
\end{equation}
By applying \eqref{A.1} twice, we can express the product of double sine functions appearing in \eqref{two-point} in Fourier space and exchange the order of integrals, performing the integrals over $s_1$ and $s_2$ first. An equivalent manipulation was performed in the case of JT gravity in \cite{Griguolo:2021zsn}. The integrals over $s_1$ and $s_2$ can be done using the following integral representation of the modified Bessel function:
\begin{equation}
K_{\nu}(z)=\int_{0}^{+\infty} e^{-z \cosh(t)} \cosh(\nu t) \mathrm{d}t.
\end{equation}
After some manipulations, the Liouville gravity two-point function \eqref{two-point} can then be rewritten as 
\begin{equation}\label{A.3}
\begin{split}
\mathcal{A}_{\beta_M}(\tau_1,\tau_2)=&\frac{C_{\Delta}}{\tau(\beta-\tau)}\int_{-\infty}^{+\infty}\mathrm{d}p \int_{-\infty}^{+\infty}\mathrm{d}q \frac{S_b(ip/b+Q/2-\beta_M)}{S_b(ip/b+Q/2+\beta_M)} \frac{S_b(iq/b+Q/2-\beta_M)}{S_b(iq/b+Q/2+\beta_M)}\\
&\bigg(q^2-\left(p-i\right)^2\bigg) \ K_{\frac{1}{b^2}\left(1+i(p+q)\right)}(\kappa \tau)K_{\frac{1}{b^2}\left(1+i(p-q)\right)}(\kappa (\beta-\tau)),
\end{split}
\end{equation}
where we set $\beta_M=\Delta b$ and the constant $C_{\Delta}=\frac{ S_b(2 \Delta b)}{b^4 (\pi b^2)^2 \kappa^2 }$.

A remarkable simplification of the above formula happens when $\Delta \in \tfrac{1}{2} \mathbb{N}$. For instance, using the property $S_b(x+b)=2 \sin (\pi b x )S_b(x)$ iteratively, the ratio of $S_{b}$ functions reduces to
\begin{equation}\label{A.4}
\frac{S_b(ip/b+Q/2-\beta_M)}{S_b(ip/b+Q/2+\beta_M)}=\frac{1}{2^{2\Delta} \cosh (\pi p)\prod_{j=1}^{\Delta-\frac12}\cosh \left(\pi  p \pm i\pi b^2 j\right)}
\end{equation}
when $\Delta$ is an half integer, and to
\begin{equation}\label{A.5}
\frac{S_b(ip/b+Q/2-\beta_M)}{S_b(ip/b+Q/2+\beta_M)}=\frac{1}{2^{2\Delta}\prod_{k=0}^{\Delta-1}\cosh \left(\pi  p\pm i\pi b^2 (k+1/2)\right)}
\end{equation}
when $\Delta$ is an integer, where $j$ and $k$ in \eqref{A.4} and \eqref{A.5} run over integer values.

Moreover, when $\Delta \in \tfrac{1}{2} \mathbb{N}$, one of the two integrals in \eqref{A.3} can be computed in closed form. Indeed, one can shift the contour for $p$ from the real line to the line at $p+i$ in the complex $p$ plane and notice the second line in \eqref{A.3} is antisymmetric in the exchange $p\leftrightarrow q$. On the other hand, one has on the first line
\begin{equation}
\frac{S_b(ip/b+Q/2-\beta_M)}{S_b(ip/b+Q/2+\beta_M)}\rightarrow \frac{S_b(ip/b+Q/2-\beta_M)}{S_b(ip/b+Q/2+\beta_M)} \times \frac{\cos \left(\pi ip/b^2+\pi/2b^2+\pi h\right)}{\cos \left(\pi ip/b^2+\pi/2b^2- \pi h\right)},
\end{equation}
which indeed is invariant, up to a factor of $(-1)^{\Delta}$, when $\Delta \in \tfrac{1}{2} \mathbb{N}$. One can check this also using the explicit forms in \eqref{A.4} and \eqref{A.5}. This property implies the full integrand in \eqref{A.3} is antisymmetric in $p\leftrightarrow q$ on the shifted contour and therefore the integral vanishes on that contour. By applying Cauchy's theorem, the integral over $p$ can then be evaluated by summing the residues of the poles located within the rectangular region bounded by the lines $\Re(p) = 0$ and $\Re(p) = i$.

When $\Delta=1/2$, we just get the residue at $p=i/2$, yielding:
\begin{equation}\label{h=1/2}
\begin{split}
\mathcal{A}_{b/2}(\tau_1,\tau_2)=&\frac{1}{b^4}\frac{S_b( b)}{(8 \pi b^2)^2 \kappa^2 \tau(\beta-\tau)} \int_{-\infty}^{+\infty}  \mathrm{d}q 
\ \frac{1+4q^2}{\cosh \left(\pi  q\right)}  \ K_{\frac{1}{b^2}\left(\frac12+iq\right)}(\kappa \tau)K_{\frac{1}{b^2}\left(\frac12-iq\right)}(\kappa (\beta-\tau)),
\end{split}
\end{equation}
so the original integral has simplified enormously. 
A possible rewriting of this formula can also be achieved using the formula:
\begin{equation}\label{Gradshteyn}
\int_{-\infty}^{+\infty} e^{i \rho x} \ K_{\nu+ix}(\alpha)K_{\nu-ix}(\beta) \mathrm{d}x=\pi \left(\frac{\alpha e^{\rho}+\beta}{\alpha+\beta e^{\rho}}\right)^{\nu} K_{2\nu}\left(\sqrt{\alpha^2+\beta^2+2\alpha \beta \cosh (\rho)}\right).
\end{equation}
In fact one can write 
\begin{equation}
\left(4 q^2+1\right) \text{sech}(\pi  q)=\frac{1}{\pi}\int_{-\infty}^{+\infty} \mathrm{d}x \ \text{sech}^3\left(\frac{x}{2}\right)\ e^{i q x},
\end{equation}
exchange the integrals and use \eqref{Gradshteyn} to get
\begin{equation}\label{h=1/2}
\begin{split}
\mathcal{A}_{b/2}(\tau_1,\tau_2)=&\frac{1}{b^4}\frac{S_b( b)}{(8 \pi b^2)^2 \kappa^2 \tau(\beta-\tau)} \int_{-\infty}^{+\infty}  \mathrm{d}x 
\ \frac{1}{\cosh^3 \left(\frac{x}{2b^2}\right)}  \ \left(\frac{\tau e^{x}+\beta-\tau}{\tau+(\beta-\tau)e^{x}}\right)^{\frac{1}{2b^2}} \times \\
&\times K_{\frac{1}{b^2}}\left(\kappa \sqrt{\tau^2+(\beta-\tau)^2+2 \tau(\beta-\tau)\cosh(x)}\right).
\end{split}
\end{equation}
Similar expressions can be obtained for $\Delta=1,3/2,2,\cdots$, where one just needs to pick more residues for evaluating the integral over $p$, since according to \eqref{A.4} and \eqref{A.5} there are more poles located in the rectangular region bounded by the lines $\Re(p) = 0$ and $\Re(p) = i$.

\bibliographystyle{ourbst}
\bibliography{Ramanujan.bib}

@article{Teschner:2012em,
    author = {Teschner, J\"org and Vartanov, Grigory},
    title = "{6j symbols for the modular double, quantum hyperbolic geometry, and supersymmetric gauge theories}",
    eprint = "1202.4698",
    archivePrefix = "arXiv",
    primaryClass = "hep-th",
    reportNumber = "DESY-12-035",
    doi = "10.1007/s11005-014-0684-3",
    journal = "Lett. Math. Phys.",
    volume = "104",
    pages = "527--551",
    year = "2014"
}

@article{Matsuzaki:1991vu,
    author = "Matsuzaki, T. and Suzuki, T.",
    title = "{A Representation of U$_q(su(1,1))$ on the space of quasiprimary fields and correlation functions}",
    reportNumber = "UTHEP-225",
    doi = "10.1016/0370-2693(92)90800-J",
    journal = "Phys. Lett. B",
    volume = "296",
    pages = "33--39",
    year = "1992"
}

@article{Shiraishi:1995rp,
    author = "Shiraishi, Jun'ichi and Kubo, Harunobu and Awata, Hidetoshi and Odake, Satoru",
    title = "{A Quantum deformation of the Virasoro algebra and the Macdonald symmetric functions}",
    eprint = "q-alg/9507034",
    archivePrefix = "arXiv",
    reportNumber = "YITP-U-95-30, DPSU-95-5, UT-715",
    doi = "10.1007/BF00398297",
    journal = "Lett. Math. Phys.",
    volume = "38",
    pages = "33--51",
    year = "1996"
}

@article{Awata:1995zk,
    author = "Awata, Hidetoshi and Kubo, Harunobu and Odake, Satoru and Shiraishi, Junichi",
    title = "{Quantum W(N) algebras and Macdonald polynomials}",
    eprint = "q-alg/9508011",
    archivePrefix = "arXiv",
    reportNumber = "YITP-U-95-34, DPSU-95-9, UT-718",
    doi = "10.1007/BF02102595",
    journal = "Commun. Math. Phys.",
    volume = "179",
    pages = "401--416",
    year = "1996"
}

@article{Nieri:2013yra,
    author = "Nieri, Fabrizio and Pasquetti, Sara and Passerini, Filippo",
    title = "{3d and 5d Gauge Theory Partition Functions as $q$-deformed CFT Correlators}",
    eprint = "1303.2626",
    archivePrefix = "arXiv",
    primaryClass = "hep-th",
    reportNumber = "DMUS-MP-13-07, CERN-PH-TH-2013-046, DMUS-MP-13/07, CERN-PH-TH/2013-046",
    doi = "10.1007/s11005-014-0727-9",
    journal = "Lett. Math. Phys.",
    volume = "105",
    number = "1",
    pages = "109--148",
    year = "2015"
}

@article{Nieri:2013vba,
    author = "Nieri, Fabrizio and Pasquetti, Sara and Passerini, Filippo and Torrielli, Alessandro",
    title = "{5D partition functions, q-Virasoro systems and integrable spin-chains}",
    eprint = "1312.1294",
    archivePrefix = "arXiv",
    primaryClass = "hep-th",
    reportNumber = "DMUS-MP-13-19, PUPT-2456, LPTENS-13-23",
    doi = "10.1007/JHEP12(2014)040",
    journal = "JHEP",
    volume = "12",
    pages = "040",
    year = "2014"
}

@article{Nedelin:2016gwu,
    author = "Nedelin, Anton and Nieri, Fabrizio and Zabzine, Maxim",
    title = "{$q$-Virasoro modular double and 3d partition functions}",
    eprint = "1605.07029",
    archivePrefix = "arXiv",
    primaryClass = "hep-th",
    reportNumber = "UUITP-10-16",
    doi = "10.1007/s00220-017-2882-1",
    journal = "Commun. Math. Phys.",
    volume = "353",
    number = "3",
    pages = "1059--1102",
    year = "2017"
}

@article{Blommaert:2025avl,
    author = "Blommaert, Andreas and Levine, Adam and Mertens, Thomas G. and Papalini, Jacopo and Parmentier, Klaas",
    title = "{Wormholes, branes and finite matrices in sine dilaton gravity}",
    eprint = "2501.17091",
    archivePrefix = "arXiv",
    primaryClass = "hep-th",
    month = "1",
    year = "2025"
}

@article{Blommaert:2023wad,
    author = "Blommaert, Andreas and Mertens, Thomas G. and Yao, Shunyu",
    title = "{The q-Schwarzian and Liouville gravity}",
    eprint = "2312.00871",
    archivePrefix = "arXiv",
    primaryClass = "hep-th",
    doi = "10.1007/JHEP11(2024)054",
    journal = "JHEP",
    volume = "11",
    pages = "054",
    year = "2024"
}

@article{Kyono:2017pxs,
    author = "Kyono, Hideki and Okumura, Suguru and Yoshida, Kentaroh",
    title = "{Comments on 2D dilaton gravity system with a hyperbolic dilaton potential}",
    eprint = "1704.07410",
    archivePrefix = "arXiv",
    primaryClass = "hep-th",
    reportNumber = "KUNS-2672",
    doi = "10.1016/j.nuclphysb.2017.07.013",
    journal = "Nucl. Phys. B",
    volume = "923",
    pages = "126--143",
    year = "2017"
}

@article{Kharchev:2001rs,
    author = "Kharchev, S. and Lebedev, D. and Semenov-Tian-Shansky, M.",
    title = "{Unitary representations of $U_q (\mathfrak{sl}(2, \mathbb{R}))$, the modular double, and the multiparticle q deformed Toda chains}",
    eprint = "hep-th/0102180",
    archivePrefix = "arXiv",
    reportNumber = "ITEP-TH-8-01",
    doi = "10.1007/s002200100592",
    journal = "Commun. Math. Phys.",
    volume = "225",
    pages = "573--609",
    year = "2002"
}

@article{Ponsot:2000mt,
    author = "Ponsot, B. and Teschner, J.",
    title = "{Clebsch-Gordan and Racah-Wigner coefficients for a continuous series of representations of U(q)(sl(2,R))}",
    eprint = "math/0007097",
    archivePrefix = "arXiv",
    reportNumber = "DIAS-STP-00-15, BERLIN-SFB-288, LPM-00-21",
    doi = "10.1007/PL00005590",
    journal = "Commun. Math. Phys.",
    volume = "224",
    pages = "613--655",
    year = "2001"
}

@article{Berkooz:2018qkz,
    author = "Berkooz, Micha and Narayan, Prithvi and Simon, Joan",
    title = "{Chord diagrams, exact correlators in spin glasses and black hole bulk reconstruction}",
    eprint = "1806.04380",
    archivePrefix = "arXiv",
    primaryClass = "hep-th",
    doi = "10.1007/JHEP08(2018)192",
    journal = "JHEP",
    volume = "08",
    pages = "192",
    year = "2018"
}

@article{Narovlansky:2023lfz,
    author = "Narovlansky, Vladimir and Verlinde, Herman",
    title = "{Double-scaled SYK and de Sitter Holography}",
    eprint = "2310.16994",
    archivePrefix = "arXiv",
    primaryClass = "hep-th",
    month = "10",
    year = "2023"
}

@article{Verlinde:2024zrh,
    author = "Verlinde, Herman and Zhang, Mengyang",
    title = "{SYK Correlators from 2D Liouville-de Sitter Gravity}",
    eprint = "2402.02584",
    archivePrefix = "arXiv",
    primaryClass = "hep-th",
    month = "2",
    year = "2024"
}

@article{Verlinde:2024znh,
    author = "Verlinde, Herman",
    title = "{Double-scaled SYK, Chords and de Sitter Gravity}",
    eprint = "2402.00635",
    archivePrefix = "arXiv",
    primaryClass = "hep-th",
    month = "2",
    year = "2024"
}

@article{Kruthoff:2024gxc,
    author = "Kruthoff, Jorrit and Levine, Adam",
    title = "{Semi-classical dilaton gravity and the very blunt defect expansion}",
    eprint = "2402.10162",
    archivePrefix = "arXiv",
    primaryClass = "hep-th",
    doi = "10.1007/JHEP07(2025)211",
    journal = "JHEP",
    volume = "07",
    pages = "211",
    year = "2025"
}

@article{Almheiri:2024xtw,
    author = "Almheiri, Ahmed and Goel, Akash and Hu, Xu-Yao",
    title = "{Quantum gravity of the Heisenberg algebra}",
    eprint = "2403.18333",
    archivePrefix = "arXiv",
    primaryClass = "hep-th",
    month = "3",
    year = "2024"
}

@article{Maldacena:1997re,
    author = "Maldacena, Juan Martin",
    title = "{The Large $N$ limit of superconformal field theories and supergravity}",
    eprint = "hep-th/9711200",
    archivePrefix = "arXiv",
    reportNumber = "HUTP-97-A097, HUTP-98-A097",
    doi = "10.4310/ATMP.1998.v2.n2.a1",
    journal = "Adv. Theor. Math. Phys.",
    volume = "2",
    pages = "231--252",
    year = "1998"
}

@article{Susskind:2021dfc,
    author = "Susskind, Leonard",
    title = "{Black Holes Hint towards De Sitter Matrix Theory}",
    eprint = "2109.01322",
    archivePrefix = "arXiv",
    primaryClass = "hep-th",
    doi = "10.3390/universe9080368",
    journal = "Universe",
    volume = "9",
    number = "8",
    pages = "368",
    year = "2023"
}

@article{Shaghoulian:2021cef,
    author = "Shaghoulian, Edgar",
    title = "{The central dogma and cosmological horizons}",
    eprint = "2110.13210",
    archivePrefix = "arXiv",
    primaryClass = "hep-th",
    doi = "10.1007/JHEP01(2022)132",
    journal = "JHEP",
    volume = "01",
    pages = "132",
    year = "2022"
}

@article{Shaghoulian:2022fop,
    author = "Shaghoulian, Edgar and Susskind, Leonard",
    title = "{Entanglement in De Sitter space}",
    eprint = "2201.03603",
    archivePrefix = "arXiv",
    primaryClass = "hep-th",
    doi = "10.1007/JHEP08(2022)198",
    journal = "JHEP",
    volume = "08",
    pages = "198",
    year = "2022"
}

@article{Franken:2023pni,
    author = "Franken, Victor and Partouche, Herv{\'e} and Rondeau, Fran{\c{c}}ois and Toumbas, Nicolaos",
    title = "{Bridging the static patches: de Sitter holography and entanglement}",
    eprint = "2305.12861",
    archivePrefix = "arXiv",
    primaryClass = "hep-th",
    reportNumber = "CPHT-RR018.042023",
    doi = "10.1007/JHEP08(2023)074",
    journal = "JHEP",
    volume = "08",
    pages = "074",
    year = "2023"
}

@article{Aguilar-Gutierrez:2024nau,
    author = "Aguilar-Gutierrez, Sergio E.",
    title = "{Towards complexity in de Sitter space from the doubled-scaled Sachdev-Ye-Kitaev model}",
    eprint = "2403.13186",
    archivePrefix = "arXiv",
    primaryClass = "hep-th",
    doi = "10.1007/JHEP10(2024)107",
    journal = "JHEP",
    volume = "10",
    pages = "107",
    year = "2024"
}

@article{Aguilar-Gutierrez:2025hty,
    author = "Aguilar-Gutierrez, Sergio E.",
    title = "{Symmetry sectors in chord space and relational holography in the DSSYK. Lessons from branes, wormholes, and de Sitter space}",
    eprint = "2506.21447",
    archivePrefix = "arXiv",
    primaryClass = "hep-th",
    doi = "10.1007/JHEP10(2025)044",
    journal = "JHEP",
    volume = "10",
    pages = "044",
    year = "2025"
}

@article{Collier:2024kmo,
    author = {Collier, Scott and Eberhardt, Lorenz and M{\"u}hlmann, Beatrix and Rodriguez, Victor A.},
    title = "{Complex Liouville String}",
    eprint = "2409.17246",
    archivePrefix = "arXiv",
    primaryClass = "hep-th",
    doi = "10.1103/k74n-s63l",
    journal = "Phys. Rev. Lett.",
    volume = "134",
    number = "25",
    pages = "251602",
    year = "2025"
}

@article{Collier:2025lux,
    author = {Collier, Scott and Eberhardt, Lorenz and M{\"u}hlmann, Beatrix},
    title = "{A microscopic realization of dS$_3$}",
    eprint = "2501.01486",
    archivePrefix = "arXiv",
    primaryClass = "hep-th",
    doi = "10.21468/SciPostPhys.18.4.131",
    journal = "SciPost Phys.",
    volume = "18",
    number = "4",
    pages = "131",
    year = "2025"
}

@article{Collier:2023cyw,
    author = {Collier, Scott and Eberhardt, Lorenz and M\"uhlmann, Beatrix and Rodriguez, Victor A.},
    title = "{The Virasoro minimal string}",
    eprint = "2309.10846",
    archivePrefix = "arXiv",
    primaryClass = "hep-th",
    doi = "10.21468/SciPostPhys.16.2.057",
    journal = "SciPost Phys.",
    volume = "16",
    number = "2",
    pages = "057",
    year = "2024"
}

@article{Hadasz:2013bwa,
    author = "Hadasz, Leszek and Pawelkiewicz, Michal and Schomerus, Volker",
    title = "{Self-dual Continuous Series of Representations for $\mathcal{U}_q(sl(2))$ and $\mathcal{U}_q(osp(1|2))$}",
    eprint = "1305.4596",
    archivePrefix = "arXiv",
    primaryClass = "hep-th",
    reportNumber = "DESY-13-083",
    doi = "10.1007/JHEP10(2014)091",
    journal = "JHEP",
    volume = "10",
    pages = "091",
    year = "2014"
}

@article{Berkooz:2018jqr,
    author = "Berkooz, Micha and Isachenkov, Mikhail and Narovlansky, Vladimir and Torrents, Genis",
    title = "{Towards a full solution of the large N double-scaled SYK model}",
    eprint = "1811.02584",
    archivePrefix = "arXiv",
    primaryClass = "hep-th",
    doi = "10.1007/JHEP03(2019)079",
    journal = "JHEP",
    volume = "03",
    pages = "079",
    year = "2019"
}

@article{Berkooz:2022mfk,
    author = "Berkooz, Micha and Isachenkov, Misha and Isachenkov, Mikhail and Narayan, Prithvi and Narovlansky, Vladimir",
    title = "{Quantum groups, non-commutative AdS$_{2}$, and chords in the double-scaled SYK model}",
    eprint = "2212.13668",
    archivePrefix = "arXiv",
    primaryClass = "hep-th",
    doi = "10.1007/JHEP08(2023)076",
    journal = "JHEP",
    volume = "08",
    pages = "076",
    year = "2023"
}

@article{Blommaert:2023opb,
    author = "Blommaert, Andreas and Mertens, Thomas G. and Yao, Shunyu",
    title = "{Dynamical actions and q-representation theory for double-scaled SYK}",
    eprint = "2306.00941",
    archivePrefix = "arXiv",
    primaryClass = "hep-th",
    month = "6",
    year = "2023"
}

@article{Polyakov:1981rd,
    author = "Polyakov, Alexander M.",
    editor = "Khalatnikov, I. M. and Mineev, V. P.",
    title = "{Quantum Geometry of Bosonic Strings}",
    reportNumber = "Print-81-0351 (LANDAU INST)",
    doi = "10.1016/0370-2693(81)90743-7",
    journal = "Phys. Lett. B",
    volume = "103",
    pages = "207--210",
    year = "1981"
}

@article{Distler:1988jt,
    author = "Distler, Jacques and Kawai, Hikaru",
    title = "{Conformal Field Theory and 2D Quantum Gravity}",
    reportNumber = "CLNS-88-853",
    doi = "10.1016/0550-3213(89)90354-4",
    journal = "Nucl. Phys. B",
    volume = "321",
    pages = "509--527",
    year = "1989"
}

@article{David:1988hj,
    author = "David, F.",
    title = "{Conformal Field Theories Coupled to 2D Gravity in the Conformal Gauge}",
    reportNumber = "SACLAY-SPHT-88-132",
    doi = "10.1142/S0217732388001975",
    journal = "Mod. Phys. Lett. A",
    volume = "3",
    pages = "1651",
    year = "1988"
}

@article{Knizhnik:1988ak,
    author = "Knizhnik, V. G. and Polyakov, Alexander M. and Zamolodchikov, A. B.",
    editor = "Khalatnikov, I. M. and Mineev, V. P.",
    title = "{Fractal Structure of 2D Quantum Gravity}",
    reportNumber = "PRINT-88-0812 (LANDAU)",
    doi = "10.1142/S0217732388000982",
    journal = "Mod. Phys. Lett. A",
    volume = "3",
    pages = "819",
    year = "1988"
}

@article{Mertens:2018fds,
    author = "Mertens, Thomas G.",
    title = "{The Schwarzian theory \textemdash{} origins}",
    eprint = "1801.09605",
    archivePrefix = "arXiv",
    primaryClass = "hep-th",
    doi = "10.1007/JHEP05(2018)036",
    journal = "JHEP",
    volume = "05",
    pages = "036",
    year = "2018"
}

@article{Jackiw:1984je,
    author = "Jackiw, R.",
    editor = "Baier, R. and Satz, H.",
    title = "{Lower Dimensional Gravity}",
    reportNumber = "MIT-CTP-1203",
    doi = "10.1016/0550-3213(85)90448-1",
    journal = "Nucl. Phys. B",
    volume = "252",
    pages = "343--356",
    year = "1985"
}

@article{Teitelboim:1983ux,
    author = "Teitelboim, C.",
    title = "{Gravitation and Hamiltonian Structure in Two Space-Time Dimensions}",
    doi = "10.1016/0370-2693(83)90012-6",
    journal = "Phys. Lett. B",
    volume = "126",
    pages = "41--45",
    year = "1983"
}

@article{Stanford:2017thb,
    author = "Stanford, Douglas and Witten, Edward",
    title = "{Fermionic Localization of the Schwarzian Theory}",
    eprint = "1703.04612",
    archivePrefix = "arXiv",
    primaryClass = "hep-th",
    doi = "10.1007/JHEP10(2017)008",
    journal = "JHEP",
    volume = "10",
    pages = "008",
    year = "2017"
}

@article{Kitaev:2018wpr,
    author = "Kitaev, Alexei and Suh, S. Josephine",
    title = "{Statistical mechanics of a two-dimensional black hole}",
    eprint = "1808.07032",
    archivePrefix = "arXiv",
    primaryClass = "hep-th",
    doi = "10.1007/JHEP05(2019)198",
    journal = "JHEP",
    volume = "05",
    pages = "198",
    year = "2019"
}

@article{Lam:2018pvp,
    author = "Lam, Ho Tat and Mertens, Thomas G. and Turiaci, Gustavo J. and Verlinde, Herman",
    title = "{Shockwave S-matrix from Schwarzian Quantum Mechanics}",
    eprint = "1804.09834",
    archivePrefix = "arXiv",
    primaryClass = "hep-th",
    doi = "10.1007/JHEP11(2018)182",
    journal = "JHEP",
    volume = "11",
    pages = "182",
    year = "2018"
}

@article{Saad:2019lba,
    author = "Saad, Phil and Shenker, Stephen H. and Stanford, Douglas",
    title = "{JT gravity as a matrix integral}",
    eprint = "1903.11115",
    archivePrefix = "arXiv",
    primaryClass = "hep-th",
    month = "3",
    year = "2019"
}

@article{Sachdev:1992fk,
    author = "Sachdev, Subir and Ye, Jinwu",
    title = "{Gapless spin fluid ground state in a random, quantum Heisenberg magnet}",
    eprint = "cond-mat/9212030",
    archivePrefix = "arXiv",
    reportNumber = "PRINT-93-0077",
    doi = "10.1103/PhysRevLett.70.3339",
    journal = "Phys. Rev. Lett.",
    volume = "70",
    pages = "3339",
    year = "1993"
}

@misc{kitaev2015talk1,
  author       = {Alexei Kitaev},
  title        = {A simple model of quantum holography (Part 1)},
  howpublished = {Talk given at the Kavli Institute for Theoretical Physics, Santa Barbara},
  month        = {April},
  year         = {2015},
  url          = {https://ucsb.edu}
}

@misc{kitaev2015talk2,
  author       = {Alexei Kitaev},
  title        = {A simple model of quantum holography (Part 2)},
  howpublished = {Talk given at the Kavli Institute for Theoretical Physics, Santa Barbara},
  month        = {May},
  year         = {2015},
  url          = {https://ucsb.edu}
}

@article{Mertens:2020hbs,
    author = "Mertens, Thomas G. and Turiaci, Gustavo J.",
    title = "{Liouville quantum gravity -- holography, JT and matrices}",
    eprint = "2006.07072",
    archivePrefix = "arXiv",
    primaryClass = "hep-th",
    doi = "10.1007/JHEP01(2021)073",
    journal = "JHEP",
    volume = "01",
    pages = "073",
    year = "2021"
}

@article{Blommaert:2024whf,
    author = "Blommaert, Andreas and Levine, Adam and Mertens, Thomas G. and Papalini, Jacopo and Parmentier, Klaas",
    title = "{An entropic puzzle in periodic dilaton gravity and DSSYK}",
    eprint = "2411.16922",
    archivePrefix = "arXiv",
    primaryClass = "hep-th",
    doi = "10.1007/JHEP07(2025)093",
    journal = "JHEP",
    volume = "07",
    pages = "093",
    year = "2025"
}

@article{Mertens:2020pfe,
    author = "Mertens, Thomas G.",
    title = "{Degenerate operators in JT and Liouville (super)gravity}",
    eprint = "2007.00998",
    archivePrefix = "arXiv",
    primaryClass = "hep-th",
    doi = "10.1007/JHEP04(2021)245",
    journal = "JHEP",
    volume = "04",
    pages = "245",
    year = "2021"
}

@article{Tietto:2025oxn,
    author = "Tietto, Damiano and Verlinde, Herman",
    title = "{A microscopic model of de Sitter spacetime with an observer}",
    eprint = "2502.03869",
    archivePrefix = "arXiv",
    primaryClass = "hep-th",
    month = "2",
    year = "2025"
}

@article{Susskind:1998dq,
    author = "Susskind, Leonard and Witten, Edward",
    title = "{The Holographic bound in anti-de Sitter space}",
    eprint = "hep-th/9805114",
    archivePrefix = "arXiv",
    reportNumber = "SU-ITP-98-39, IASSNS-HEP-98-44",
    month = "5",
    year = "1998"
}

@article{Giveon:1999px,
    author = "Giveon, Amit and Kutasov, David",
    title = "{Little string theory in a double scaling limit}",
    eprint = "hep-th/9909110",
    archivePrefix = "arXiv",
    reportNumber = "NSF-ITP-99-104, RI-8-99, EFI-99-41",
    doi = "10.1088/1126-6708/1999/10/034",
    journal = "JHEP",
    volume = "10",
    pages = "034",
    year = "1999"
}

@article{Klebanov:2000hb,
    author = "Klebanov, Igor R. and Strassler, Matthew J.",
    title = "{Supergravity and a confining gauge theory: Duality cascades and chi SB resolution of naked singularities}",
    eprint = "hep-th/0007191",
    archivePrefix = "arXiv",
    reportNumber = "IASSNS-HEP-00-56, PUPT-1944",
    doi = "10.1088/1126-6708/2000/08/052",
    journal = "JHEP",
    volume = "08",
    pages = "052",
    year = "2000"
}

@article{Kanitscheider:2008kd,
    author = "Kanitscheider, Ingmar and Skenderis, Kostas and Taylor, Marika",
    title = "{Precision holography for non-conformal branes}",
    eprint = "0807.3324",
    archivePrefix = "arXiv",
    primaryClass = "hep-th",
    reportNumber = "ITFA-2008-25",
    doi = "10.1088/1126-6708/2008/09/094",
    journal = "JHEP",
    volume = "09",
    pages = "094",
    year = "2008"
}

@article{Taylor:2015glc,
    author = "Taylor, Marika",
    title = "{Lifshitz holography}",
    eprint = "1512.03554",
    archivePrefix = "arXiv",
    primaryClass = "hep-th",
    doi = "10.1088/0264-9381/33/3/033001",
    journal = "Class. Quant. Grav.",
    volume = "33",
    number = "3",
    pages = "033001",
    year = "2016"
}

@article{Nayak:2018qej,
    author = "Nayak, Pranjal and Shukla, Ashish and Soni, Ronak M. and Trivedi, Sandip P. and Vishal, V.",
    title = "{On the Dynamics of Near-Extremal Black Holes}",
    eprint = "1802.09547",
    archivePrefix = "arXiv",
    primaryClass = "hep-th",
    reportNumber = "TIFR/TH/17-35, TIFR-TH-17-35",
    doi = "10.1007/JHEP09(2018)048",
    journal = "JHEP",
    volume = "09",
    pages = "048",
    year = "2018"
}

@article{Iliesiu:2020qvm,
    author = "Iliesiu, Luca V. and Turiaci, Gustavo J.",
    title = "{The statistical mechanics of near-extremal black holes}",
    eprint = "2003.02860",
    archivePrefix = "arXiv",
    primaryClass = "hep-th",
    doi = "10.1007/JHEP05(2021)145",
    journal = "JHEP",
    volume = "05",
    pages = "145",
    year = "2021"
}

@article{Goel:2023svz,
    author = "Goel, Akash and Narovlansky, Vladimir and Verlinde, Herman",
    title = "{Semiclassical geometry in double-scaled SYK}",
    eprint = "2301.05732",
    archivePrefix = "arXiv",
    primaryClass = "hep-th",
    month = "1",
    year = "2023"
}

@article{Lin:2022rbf,
    author = "Lin, Henry W.",
    title = "{The bulk Hilbert space of double scaled SYK}",
    eprint = "2208.07032",
    archivePrefix = "arXiv",
    primaryClass = "hep-th",
    doi = "10.1007/JHEP11(2022)060",
    journal = "JHEP",
    volume = "11",
    pages = "060",
    year = "2022"
}

@article{Lin:2023trc,
    author = "Lin, Henry W. and Stanford, Douglas",
    title = "{A symmetry algebra in double-scaled SYK}",
    eprint = "2307.15725",
    archivePrefix = "arXiv",
    primaryClass = "hep-th",
    doi = "10.21468/SciPostPhys.15.6.234",
    journal = "SciPost Phys.",
    volume = "15",
    number = "6",
    pages = "234",
    year = "2023"
}

@article{Suzuki:2021zbe,
    author = "Suzuki, Kenta and Takayanagi, Tadashi",
    title = "{JT gravity limit of Liouville CFT and matrix model}",
    eprint = "2108.12096",
    archivePrefix = "arXiv",
    primaryClass = "hep-th",
    reportNumber = "YITP-21-88, IPMU21-0054",
    doi = "10.1007/JHEP11(2021)137",
    journal = "JHEP",
    volume = "11",
    pages = "137",
    year = "2021"
}

@article{Fateev:2000ik,
    author = "Fateev, V. and Zamolodchikov, Alexander B. and Zamolodchikov, Alexei B.",
    title = "{Boundary Liouville field theory. 1. Boundary state and boundary two point function}",
    eprint = "hep-th/0001012",
    archivePrefix = "arXiv",
    reportNumber = "RUNHETC-2000-01",
    month = "1",
    year = "2000"
}

@article{Teschner:2000md,
    author = "Teschner, J.",
    editor = "Bernard, Denis and Bonora, Loriano and Mussardo, Giuseppe and Corrigan, Edward and Gomez, Cesar and Nahm, Werner and Julia, Bernard",
    title = "{Remarks on Liouville theory with boundary}",
    eprint = "hep-th/0009138",
    archivePrefix = "arXiv",
    doi = "10.22323/1.006.0041",
    journal = "PoS",
    volume = "tmr2000",
    pages = "041",
    year = "2000"
}

@article{Harlow:2018tqv,
    author = "Harlow, Daniel and Jafferis, Daniel",
    title = "{The Factorization Problem in Jackiw-Teitelboim Gravity}",
    eprint = "1804.01081",
    archivePrefix = "arXiv",
    primaryClass = "hep-th",
    doi = "10.1007/JHEP02(2020)177",
    journal = "JHEP",
    volume = "02",
    pages = "177",
    year = "2020"
}

@article{Belaey:2025ijg,
    author = "Belaey, Andreas and Mertens, Thomas G. and Tappeiner, Thomas",
    title = "{Quantum group origins of edge states in double-scaled SYK}",
    eprint = "2503.20691",
    archivePrefix = "arXiv",
    primaryClass = "hep-th",
    doi = "10.1007/JHEP02(2026)184",
    journal = "JHEP",
    volume = "02",
    pages = "184",
    year = "2026"
}

@article{Almheiri:2024ayc,
    author = "Almheiri, Ahmed and Popov, Fedor K.",
    title = "{Holography on the quantum disk}",
    eprint = "2401.05575",
    archivePrefix = "arXiv",
    primaryClass = "hep-th",
    doi = "10.1007/JHEP06(2024)070",
    journal = "JHEP",
    volume = "06",
    pages = "070",
    year = "2024"
}

@article{Bernard:1989jq,
    author = "Bernard, Denis and LeClair, Andre",
    title = "{$q$ Deformation of SU(1,1) Conformal Ward Identities and $q$ Strings}",
    reportNumber = "PUPT-1123, CLNS-89/933",
    doi = "10.1016/0370-2693(89)90953-2",
    journal = "Phys. Lett. B",
    volume = "227",
    pages = "417--423",
    year = "1989"
}

@article{Witten:2021unn,
    author = "Witten, Edward",
    title = "{Gravity and the crossed product}",
    eprint = "2112.12828",
    archivePrefix = "arXiv",
    primaryClass = "hep-th",
    doi = "10.1007/JHEP10(2022)008",
    journal = "JHEP",
    volume = "10",
    pages = "008",
    year = "2022"
}

@article{Witten:2020ert,
    author = "Witten, Edward",
    title = "{Deformations of JT Gravity and Phase Transitions}",
    eprint = "2006.03494",
    archivePrefix = "arXiv",
    primaryClass = "hep-th",
    month = "6",
    year = "2020"
}

@article{Chandrasekaran:2022cip,
    author = "Chandrasekaran, Venkatesa and Longo, Roberto and Penington, Geoff and Witten, Edward",
    title = "{An algebra of observables for de Sitter space}",
    eprint = "2206.10780",
    archivePrefix = "arXiv",
    primaryClass = "hep-th",
    doi = "10.1007/JHEP02(2023)082",
    journal = "JHEP",
    volume = "02",
    pages = "082",
    year = "2023"
}

@article{Jensen:2023yxy,
    author = "Jensen, Kristan and Sorce, Jonathan and Speranza, Antony J.",
    title = "{Generalized entropy for general subregions in quantum gravity}",
    eprint = "2306.01837",
    archivePrefix = "arXiv",
    primaryClass = "hep-th",
    doi = "10.1007/JHEP12(2023)020",
    journal = "JHEP",
    volume = "12",
    pages = "020",
    year = "2023"
}

@article{Penington:2019npb,
    author = "Penington, Geoffrey",
    title = "{Entanglement Wedge Reconstruction and the Information Paradox}",
    eprint = "1905.08255",
    archivePrefix = "arXiv",
    primaryClass = "hep-th",
    doi = "10.1007/JHEP09(2020)002",
    journal = "JHEP",
    volume = "09",
    pages = "002",
    year = "2020"
}

@article{Almheiri:2019psf,
    author = "Almheiri, Ahmed and Engelhardt, Netta and Marolf, Donald and Maxfield, Henry",
    title = "{The entropy of bulk quantum fields and the entanglement wedge of an evaporating black hole}",
    eprint = "1905.08762",
    archivePrefix = "arXiv",
    primaryClass = "hep-th",
    doi = "10.1007/JHEP12(2019)063",
    journal = "JHEP",
    volume = "12",
    pages = "063",
    year = "2019"
}

@article{Hartman:2013qma,
    author = "Hartman, Thomas and Maldacena, Juan",
    title = "{Time Evolution of Entanglement Entropy from Black Hole Interiors}",
    eprint = "1303.1080",
    archivePrefix = "arXiv",
    primaryClass = "hep-th",
    doi = "10.1007/JHEP05(2013)014",
    journal = "JHEP",
    volume = "05",
    pages = "014",
    year = "2013"
}

@article{Brown:2015bva,
    author = "Brown, Adam R. and Roberts, Daniel A. and Susskind, Leonard and Swingle, Brian and Zhao, Ying",
    title = "{Holographic Complexity Equals Bulk Action?}",
    eprint = "1509.07876",
    archivePrefix = "arXiv",
    primaryClass = "hep-th",
    doi = "10.1103/PhysRevLett.116.191301",
    journal = "Phys. Rev. Lett.",
    volume = "116",
    number = "19",
    pages = "191301",
    year = "2016"
}

@article{Keeler:2014hba,
    author = "Keeler, Cynthia and Ng, Gim Seng",
    title = "{Partition Functions in Even Dimensional AdS via Quasinormal Mode Methods}",
    eprint = "1401.7016",
    archivePrefix = "arXiv",
    primaryClass = "hep-th",
    doi = "10.1007/JHEP06(2014)099",
    journal = "JHEP",
    volume = "06",
    pages = "099",
    year = "2014"
}

@article{Fan:2021bwt,
    author = "Fan, Yale and Mertens, Thomas G.",
    title = "{From quantum groups to Liouville and dilaton quantum gravity}",
    eprint = "2109.07770",
    archivePrefix = "arXiv",
    primaryClass = "hep-th",
    doi = "10.1007/JHEP05(2022)092",
    journal = "JHEP",
    volume = "05",
    pages = "092",
    year = "2022"
}

@article{Mertens:2022aou,
    author = "Mertens, Thomas G.",
    title = "{Quantum exponentials for the modular double and applications in gravity models}",
    eprint = "2212.07696",
    archivePrefix = "arXiv",
    primaryClass = "hep-th",
    doi = "10.1007/JHEP09(2023)106",
    journal = "JHEP",
    volume = "09",
    pages = "106",
    year = "2023"
}

@article{Cotler:2016fpe,
    author = "Cotler, Jordan S. and Gur-Ari, Guy and Hanada, Masanori and Polchinski, Joseph and Saad, Phil and Shenker, Stephen H. and Stanford, Douglas and Streicher, Alexandre and Tezuka, Masaki",
    title = "{Black Holes and Random Matrices}",
    eprint = "1611.04650",
    archivePrefix = "arXiv",
    primaryClass = "hep-th",
    reportNumber = "SU-ITP-16-19, SU-ITP-16/19, YITP-16-124",
    doi = "10.1007/JHEP05(2017)118",
    journal = "JHEP",
    volume = "05",
    pages = "118",
    year = "2017",
    note = "[Erratum: JHEP 09, 002 (2018)]"
}

@article{Almheiri:2014cka,
    author = "Almheiri, Ahmed and Polchinski, Joseph",
    title = "{Models of AdS$_{2}$ backreaction and holography}",
    eprint = "1402.6334",
    archivePrefix = "arXiv",
    primaryClass = "hep-th",
    doi = "10.1007/JHEP11(2015)014",
    journal = "JHEP",
    volume = "11",
    pages = "014",
    year = "2015"
}

@article{Maldacena:2016upp,
    author = "Maldacena, Juan and Stanford, Douglas and Yang, Zhenbin",
    title = "{Conformal symmetry and its breaking in two dimensional Nearly Anti-de-Sitter space}",
    eprint = "1606.01857",
    archivePrefix = "arXiv",
    primaryClass = "hep-th",
    doi = "10.1093/ptep/ptw124",
    journal = "PTEP",
    volume = "2016",
    number = "12",
    pages = "12C104",
    year = "2016"
}

@article{Jensen:2016pah,
    author = "Jensen, Kristan",
    title = "{Chaos in AdS$_2$ Holography}",
    eprint = "1605.06098",
    archivePrefix = "arXiv",
    primaryClass = "hep-th",
    doi = "10.1103/PhysRevLett.117.111601",
    journal = "Phys. Rev. Lett.",
    volume = "117",
    number = "11",
    pages = "111601",
    year = "2016"
}

@article{Engelsoy:2016xyb,
    author = {Engels\"oy, Julius and Mertens, Thomas G. and Verlinde, Herman},
    title = "{An investigation of AdS$_{2}$ backreaction and holography}",
    eprint = "1606.03438",
    archivePrefix = "arXiv",
    primaryClass = "hep-th",
    doi = "10.1007/JHEP07(2016)139",
    journal = "JHEP",
    volume = "07",
    pages = "139",
    year = "2016"
}

@article{Mertens:2017mtv,
    author = "Mertens, Thomas G. and Turiaci, Gustavo J. and Verlinde, Herman L.",
    title = "{Solving the Schwarzian via the Conformal Bootstrap}",
    eprint = "1705.08408",
    archivePrefix = "arXiv",
    primaryClass = "hep-th",
    doi = "10.1007/JHEP08(2017)136",
    journal = "JHEP",
    volume = "08",
    pages = "136",
    year = "2017"
}

@article{Mertens:2022irh,
    author = "Mertens, Thomas G. and Turiaci, Gustavo J.",
    title = "{Solvable models of quantum black holes: a review on Jackiw\textendash{}Teitelboim gravity}",
    eprint = "2210.10846",
    archivePrefix = "arXiv",
    primaryClass = "hep-th",
    doi = "10.1007/s41114-023-00046-1",
    journal = "Living Rev. Rel.",
    volume = "26",
    number = "1",
    pages = "4",
    year = "2023"
}

@article{Ponsot:1999uf,
    author = "Ponsot, B. and Teschner, J.",
    title = "{Liouville bootstrap via harmonic analysis on a noncompact quantum group}",
    eprint = "hep-th/9911110",
    archivePrefix = "arXiv",
    reportNumber = "DIAS-STP-99-14, ESI-783, LPM-99-46",
    month = "11",
    year = "1999"
}

@article{Ip:2013,
      author         = "Ivan C-H. Ip",
      title          = "{Representation of the quantum plane, its quantum double and harmonic analysis on $GL_q^+(2,R)$}",
      journal        = "Selecta Mathematica New Series",
      volume         = "Vol 19 (4)",
      year           = "2013",
      pages          = "987-1082",
      doi            = "10.1007/s00029-012-0112-4",
      eprint         = "1108.5365",
      archivePrefix  = "arXiv",
      primaryClass   = "math.QA",
}

@article{Susskind:2022bia,
    author = "Susskind, Leonard",
    title = "{De Sitter Space, Double-Scaled SYK, and the Separation of Scales in the Semiclassical Limit}",
    eprint = "2209.09999",
    archivePrefix = "arXiv",
    primaryClass = "hep-th",
    month = "9",
    year = "2022"
}

@article{Bhattacharjee:2022ave,
    author = "Bhattacharjee, Budhaditya and Nandy, Pratik and Pathak, Tanay",
    title = "{Krylov complexity in large q and double-scaled SYK model}",
    eprint = "2210.02474",
    archivePrefix = "arXiv",
    primaryClass = "hep-th",
    reportNumber = "YITP-22-106",
    doi = "10.1007/JHEP08(2023)099",
    journal = "JHEP",
    volume = "08",
    pages = "099",
    year = "2023"
}

@article{Mukhametzhanov:2023tcg,
    author = "Mukhametzhanov, Baur",
    title = "{Large p SYK from chord diagrams}",
    eprint = "2303.03474",
    archivePrefix = "arXiv",
    primaryClass = "hep-th",
    doi = "10.1007/JHEP09(2023)154",
    journal = "JHEP",
    volume = "09",
    pages = "154",
    year = "2023"
}

@article{Berkooz:2023cqc,
    author = "Berkooz, Micha and Jia, Yiyang and Silberstein, Navot",
    title = "{Parisi\textquoteright{}s Hypercube, Fock-Space Frustration, and Near-AdS2/Near-CFT1 Holography}",
    eprint = "2303.18182",
    archivePrefix = "arXiv",
    primaryClass = "hep-th",
    doi = "10.1103/PhysRevLett.132.081601",
    journal = "Phys. Rev. Lett.",
    volume = "132",
    number = "8",
    pages = "081601",
    year = "2024"
}

@article{Okuyama:2023bch,
    author = "Okuyama, Kazumi and Suzuki, Kenta",
    title = "{Correlators of double scaled SYK at one-loop}",
    eprint = "2303.07552",
    archivePrefix = "arXiv",
    primaryClass = "hep-th",
    reportNumber = "RUP-23-5",
    doi = "10.1007/JHEP05(2023)117",
    journal = "JHEP",
    volume = "05",
    pages = "117",
    year = "2023"
}

@article{Lin:2022nss,
    author = "Lin, Henry and Susskind, Leonard",
    title = "{Infinite Temperature's Not So Hot}",
    eprint = "2206.01083",
    archivePrefix = "arXiv",
    primaryClass = "hep-th",
    month = "6",
    year = "2022"
}

@article{Bossi:2024ffa,
    author = "Bossi, Leonardo and Griguolo, Luca and Papalini, Jacopo and Russo, Lorenzo and Seminara, Domenico",
    title = "{Sine-dilaton gravity vs double-scaled SYK: exploring one-loop quantum corrections}",
    eprint = "2411.15957",
    archivePrefix = "arXiv",
    primaryClass = "hep-th",
    doi = "10.1007/JHEP06(2025)152",
    journal = "JHEP",
    volume = "06",
    pages = "152",
    year = "2025"
}

@article{Xu:2024hoc,
    author = "Xu, Jiuci",
    title = "{Von Neumann algebras in double-scaled SYK}",
    eprint = "2403.09021",
    archivePrefix = "arXiv",
    primaryClass = "hep-th",
    doi = "10.1007/JHEP03(2026)016",
    journal = "JHEP",
    volume = "03",
    pages = "016",
    year = "2026"
}

@article{Xu:2024gfm,
    author = "Xu, Jiuci",
    title = "{On chord dynamics and complexity growth in double-scaled SYK}",
    eprint = "2411.04251",
    archivePrefix = "arXiv",
    primaryClass = "hep-th",
    doi = "10.1007/JHEP06(2025)259",
    journal = "JHEP",
    volume = "06",
    pages = "259",
    year = "2025"
}

@article{Susskind:2023hnj,
    author = "Susskind, Leonard",
    title = "{De Sitter Space has no Chords. Almost Everything is Confined.}",
    eprint = "2303.00792",
    archivePrefix = "arXiv",
    primaryClass = "hep-th",
    doi = "10.22128/jhap.2023.661.1043",
    journal = "JHAP",
    volume = "3",
    number = "1",
    pages = "1--30",
    year = "2023"
}

@article{Berkooz:2024evs,
    author = "Berkooz, Micha and Brukner, Nadav and Jia, Yiyang and Mamroud, Ohad",
    title = "{From Chaos to Integrability in Double Scaled Sachdev-Ye-Kitaev Model via a Chord Path Integral}",
    eprint = "2403.01950",
    archivePrefix = "arXiv",
    primaryClass = "hep-th",
    doi = "10.1103/PhysRevLett.133.221602",
    journal = "Phys. Rev. Lett.",
    volume = "133",
    number = "22",
    pages = "221602",
    year = "2024"
}

@article{Heller:2024ldz,
    author = "Heller, Michal P. and Papalini, Jacopo and Schuhmann, Tim",
    title = "{Krylov Spread Complexity as Holographic Complexity beyond Jackiw-Teitelboim Gravity}",
    eprint = "2412.17785",
    archivePrefix = "arXiv",
    primaryClass = "hep-th",
    doi = "10.1103/spcr-jgm6",
    journal = "Phys. Rev. Lett.",
    volume = "135",
    number = "15",
    pages = "151602",
    year = "2025"
}

@inproceedings{Faddeev:1999fe,
    author = "Faddeev, L. D.",
    title = "{Modular double of quantum group}",
    booktitle = "{Conference Moshe Flato}",
    eprint = "math/9912078",
    archivePrefix = "arXiv",
    pages = "149--156",
    year = "2000"
}

@article{Bytsko:2002br,
    author = "Bytsko, A. G. and Teschner, J.",
    title = "{R operator, coproduct and Haar measure for the modular double of U(q)(sl(2,R))}",
    eprint = "math/0208191",
    archivePrefix = "arXiv",
    doi = "10.1007/s00220-003-0894-5",
    journal = "Commun. Math. Phys.",
    volume = "240",
    pages = "171--196",
    year = "2003"
}

@article{Bytsko:2006ut,
    author = "Bytsko, Andrei G. and Teschner, Jorg",
    title = "{Quantization of models with non-compact quantum group symmetry: Modular XXZ magnet and lattice sinh-Gordon model}",
    eprint = "hep-th/0602093",
    archivePrefix = "arXiv",
    doi = "10.1088/0305-4470/39/41/S11",
    journal = "J. Phys. A",
    volume = "39",
    pages = "12927--12981",
    year = "2006"
}

@article{article,
author = {Kleinert, Hagen},
year = {2004},
month = {03},
pages = {},
title = {Path Integrals In Quantum Mechanics, Statistics, Polymer Physics, And Financial Markets},
isbn = {978-981-238-106-4},
journal = {Path integrals in quantum mechanics, statistics, polymer physics, and financial markets, 3rd ed., by Hagen Kleinert. Singapore: World Scientific Publishing, 2004},
doi = {10.1142/5057}
}

@article{Griguolo:2021zsn,
    author = "Griguolo, Luca and Papalini, Jacopo and Seminara, Domenico",
    title = "{On the perturbative expansion of exact bi-local correlators in JT gravity}",
    eprint = "2101.06252",
    archivePrefix = "arXiv",
    primaryClass = "hep-th",
    doi = "10.1007/JHEP05(2021)140",
    journal = "JHEP",
    volume = "05",
    pages = "140",
    year = "2021"
}

@article{Berkooz:2025ydg,
    author = "Berkooz, Micha and Kukolj, Trivko and Seitz, Josef",
    title = "{Comments on class S(YK)}",
    eprint = "2507.12524",
    archivePrefix = "arXiv",
    primaryClass = "hep-th",
    doi = "10.1007/JHEP02(2026)037",
    journal = "JHEP",
    volume = "02",
    pages = "037",
    year = "2026"
}

@article{Blommaert:2024ymv,
    author = "Blommaert, Andreas and Mertens, Thomas G. and Papalini, Jacopo",
    title = "{The dilaton gravity hologram of double-scaled SYK}",
    eprint = "2404.03535",
    archivePrefix = "arXiv",
    primaryClass = "hep-th",
    doi = "10.1007/JHEP06(2025)050",
    journal = "JHEP",
    volume = "06",
    pages = "050",
    year = "2025"
}
\end{document}